\documentclass[fleqn,12pt]{article}
\usepackage{graphics,epsfig,here,mathrsfs}

\def\beq   {\begin{equation}}
\def\eeq   {\end{equation}}
\def\beqd  {\begin{displaymath}}
\def\eeqd  {\end{displaymath}}
\def\beqaa {\begin{eqnarray}}
\def\eeqaa {\end{eqnarray}}

\def\PR              {P_R^{}}
\def\PL              {P_L^{}}

\def\Rst             {{\cal R}^{\st}}

\def\tW              {\t_W}

\def\rzw             {\sqrt{2}}

\def\noi {\noindent}

\def\ti  {\tilde}

\def\st  {\ti t}

\def\nt  {\tilde\chi^0}

\def\b   {\beta}
\def\t   {\theta}

\def\sz{\ifmmode{\tilde{\chi}^0} \else{$\tilde{\chi}^0$} \fi}
\def\sw{\ifmmode{\tilde{\chi}} \else{$\tilde{\chi}$} \fi}

\newcommand{\gsim}{\;\raisebox{-0.9ex}
           {$\textstyle\stackrel{\textstyle >}{\sim}$}\;}
\newcommand{\lsim}{\;\raisebox{-0.9ex}{$\textstyle\stackrel{\textstyle<}
           {\sim}$}\;}

\newcommand{\be}[1]{\begin{equation} \label{(#1)}}
\newcommand{\ee}{\end{equation}}
\newcommand{\baq}[1]{\begin{eqnarray} \label{(#1)}}
\newcommand{\eaq}{\end{eqnarray}}
\newcommand{\rf}[1]{(\ref{(#1)})}
\newcommand{\ba}{\begin{array}}
\newcommand{\ea}{\end{array}}

\newcommand{\slashed}[1]{\not\!#1}

\begin{document}

\vspace*{-1cm}
\begin{flushright}
  UWThPh-2004-20 \\
\end{flushright}

\vspace*{1.4cm}

\begin{center}

{\Large {\bf Triple product correlations in top squark decays
}}\\

\vspace{2cm}

{\large
A.~Bartl$^a$, E.~Christova$^b$, K.~Hohenwarter-Sodek$^a$, T.~Kernreiter$^a$}

\vspace{2cm}

$^a${\it Institut f\"ur Theoretische Physik, Universit\"at Wien, A-1090
Vienna, Austria}
$^b${\it Institute for Nuclear Research and Nuclear Energy,
Sofia 1784, Bulgaria}

\end{center}

\vspace{3cm}

\begin{abstract}

We propose several T-odd asymmetries in the decay chains
of the top squarks $\tilde t_m \to t \tilde \chi^0_k$
and $t\to bW^+\to bl\nu$ and
$\tilde\chi^0_k \to l^\pm\tilde l_n^\mp \to l^\pm l^\mp\tilde\chi^0_1$,
for $l =e,\mu,\tau$. We calculate the asymmetries within the
Minimal Supersymmetric Standard Model with complex parameters 
$M_1$, $\mu$ and $A_t$. We give the analytic formulae for the decay 
distributions. We present numerical results for the asymmetries
and estimate the event rates necessary to observe them. 
The largest T-odd asymmetry can be as large as 40\%.

\end{abstract}




\newpage


\section{Introduction}

In the Minimal Supersymmetric Standard Model (MSSM) \cite{susy,guha} 
with complex parameters, there are new sources of CP violation 
in addition to the CKM phase of the Standard Model (SM).
After redefining the fields these  are the
phase of the higgsino mass parameter $\mu$,  two of the phases of the
gaugino masses $M_{i}$, $i = 1,2,3$ (usually these are chosen to
be the phases of $M_1$ and $M_3$), and the phases of the trilinear
couplings $A_{f}$, $\phi_{A_{f}}$. 
The latter ones for the third generations $f=\tau,t,b$ 
are rather unconstrained by the experimental upper bounds on the electric
dipole moments of electron and neutron \cite{Pilaftsis}.
It is therefore especially interesting to search for observables
which could be probed in forthcoming collider experiments in order to
determine the phases $\phi_{A_{\tau,t,b}}$.
The influence of the phases $\phi_{A_{\tau,t,b}}$ has been 
discussed in the literature before.
Some examples of studies discussing CP sensitive observables 
in SM processes or in processes which might occur 
in the SM with an extended Higgs sector are in \cite{Bernreuther:1995nw}, 
where the influence of the phases $\phi_{A_{t,b}}$ arises
due to loop corrections. Other studies focus on the $\phi_{A_{\tau,t,b}}$ 
dependence in supersymmetric processes.
There the dependence of $\phi_{A_{\tau,t,b}}$ on either 
CP-odd observables \cite{Choi:1998ub}
or on CP-even observables \cite{Bartl:2002bh,Bartl:2003pd,Bartl:2002uy} have been 
discussed.
A CP sensitive asymmetry in the 3-body decay 
$\ti t_1\to b \ti\nu_{\tau}\tau^+$ involving the transverse
polarization of the $\tau$ lepton has been proposed in \cite{Bartl:2002hi}.

In this paper we investigate whether the search for aplanarities in
the decay chain of the top squarks $\ti t_{1,2}$
can give information on the CP phase $\phi_{A_t}$
or on other couplings of the MSSM Lagrangian.
We consider the decay chain
\begin{eqnarray}
\tilde t_m \to t \tilde \chi^0_k, \label{st}
\end{eqnarray}
with the subsequent decays of $t$ and $\tilde\chi^0_k$.
We work in the approximation when both
the top quark and the neutralino $\tilde\chi^0_k$ are produced
on mass-shell.
As the top quark does not form a bound state
(because of its large mass),  both the top quark and 
the neutralino $\tilde\chi^0_k$
decay with definite momentum and polarization.
We consider two possibilities for the top quark decay:
\begin{eqnarray}
t\to bW^+\qquad {\rm and}\qquad
t\to bW^+\to b l\nu (b c s),
\label{t}
\end{eqnarray}
and the following two-decay chains for $\tilde\chi^0_k$:
\begin{eqnarray}
\tilde\chi^0_k \to \tilde l_n^-l_1^+,\quad
\tilde l_n^- \to l_2^-\tilde\chi^0_1
\qquad {\rm and}\qquad
\tilde\chi^0_k \to \tilde l_n^+l_1^-,\quad
\tilde l_n^+ \to l_2^+\tilde\chi^0_1~,
\label{chik}
\end{eqnarray}
where the label of the leptons indicates their origin and where
both $l^\pm_1$ and $l^\mp_2$ are from the same lepton family. (Sometimes in literature
$l^\pm_1$ and $l^\mp_2$ are called the near and far lepton, see e.g. 
\cite{Barr:2004ze}.)
We assume that the momenta of all ordinary particles in (\ref{st})-(\ref{chik})
can be measured or reconstructed, these are $p_t$, $p_b$, $p_l$, 
$p_{l_1^\pm}$
and $p_{l_2^\mp}$.
The final state consists of two opposite signed leptons of the same
family, $l^+ l^-$, a $b$ quark and $q \bar q'$ jets (or $l$) 
from the $t$ quark decay and missing energy.

An useful tool for studying CP violation  are triple product correlations
$({\mathbf q}_1\times {\mathbf q}_2\cdot \mathbf q_3)$ $\equiv$
$({\mathbf q}_1 {\mathbf q}_2 {\mathbf q}_3)$ 
\cite{Bernreuther:1989kc,Valencia:1994zi},
where ${\mathbf q}_i$ can be any of the 3-momenta of the particles in the decay chain.
Triple product correlations are an example of T-odd correlations
that  change sign under a flip of the 3-momenta ${\mathbf q}_i \to -\,{\mathbf q}_i$.
The time reversal operation T implies  not only reverse of the 3-momenta
and polarizations of the particles but also an
interchange of the initial and final states. Because of the
antiunitary nature of the time reversal operation, a non-zero
value of a T-odd observable would imply T-violation if
loop amplitudes are neglected. 
Any triple product correlation would be a direct evidence
that T invariance is broken and as
CPT invariance holds, CP conservation as well. As the
triple product correlations in the processes (\ref{st})-(\ref{chik}) 
are a tree-level effect, they do not contain the  suppression factor
due to radiative corrections that is always present when such
correlations are considered in processes with ordinary particles.

In the top squark decays (\ref{st})-(\ref{chik})
no triple product correlations can arise solely
from the decays of either $t$ or $\tilde\chi^0_k$. Triple products
originate from the covariant products $\varepsilon (q_1q_2q_3q_4)$
written in the laboratory system. In order that 
$\varepsilon(q_1q_2q_3q_4)\neq 0$ leads to a CP asymmetry at tree-level 
we need both a CP violating
phase and at least a 3-body decay mediated by a particle that
is not a scalar. The top quark decay modes (\ref{t}) proceed in
the SM and  at tree-level no CP violating phases occur, thus no
correlations of the type $(\mathbf{p}_l\mathbf{ p}_b
\mathbf{p}_t)$ can appear.  The $\ti\chi^0_k$ decays (\ref{chik}) are 3-body
decays mediated by the scalar lepton $\tilde l_n$ and, as
$\tilde l_n$ does not transfer information about the spin of $\tilde\chi^0_k$
to its decay products,
 again no triple products can be formed.

Thus, the only correlations which occur are among the momenta
of the decay products of both $t$ and $\tilde\chi^0_k$. 
These correlations reflect the spin properties of $t$ and $\ti\chi^0_k$.
In order to obtain analytic expressions for the distributions
of the decay products we use the
formalism of Kawasaki, Shirafuji and Tsai \cite{K&T}. We work in
the narrow width approximation for
$t$ and $\tilde\chi^0_k$.

As T-odd observables we consider up-down asymmetries,
which are defined by
\be{eq:cpasy}
A_T\equiv\frac{\int d\Omega ~sgn({\mathcal O}) ~d\Gamma/d\Omega}
{\int d\Omega ~d\Gamma/d\Omega} =
\frac{N[{\mathcal O}> 0]-N[{\mathcal O} < 0]}
{N[{\mathcal O}> 0]+N[{\mathcal O} < 0]},
\ee
where $d\Gamma$ stands for the differential decay width and
$d\Omega$ involves the angles of integration.
In Eq.~\rf{eq:cpasy} ${\mathcal O}$ represents the triple product correlation
on which we focus and $N[{\mathcal O} > (<)~ 0]$ is the number
of events for which ${\mathcal O}> (<) ~0$.
According to the decay channels of the top quark we consider two cases:

1)
If  $t \to bW$, the possible triple products are
\be{eq:triplehad}
({\bf p}_b {\bf p}_t {\bf l}_{1,2}^\pm )
\quad {\rm and}\quad ({\bf p}_b {\bf l}_1^+ {\bf l}_2^-).
\ee

2) If a final  leptonic ($\nu l$) or  hadronic ($c s$) decay mode of W  is 
measured, then possible triple product correlations are:
\be{eq:triplelep}
({\bf p}_t {\bf p}_{l,c} {\bf l}_{1,2}^\pm )~,\quad
({\bf p}_b {\bf p}_{l,c} {\bf l}_{1,2}^\pm )
\quad {\rm and}\quad
({\bf p}_{l,c} {\bf l}_1^+ {\bf l}_2^-)~.
\ee
In most of the asymmetries studied below $b$-tagging will be
necessary. In those asymmetries which involve the decay $W\to c s$
also $c$-tagging will be necessary \cite{Damerell:1996sv}.

The decay $\ti\chi^0_k \to Z^0 \tilde \chi^0_1, Z^0 \to l^+l^-$,
leads to the same final state as decay (\ref{chik}) 
and gives also rise to the above triple product correlations. 
In this paper we will not consider triple product correlations in
this decay, because due to the small $Z^0l^+l^-$ vector coupling 
one can expect that the corresponding T-odd asymmetries are
much smaller than those following from the decay (\ref{chik}).
(In this context see also \cite{Choi:1999cc,Bartl:2003ck,Bartl:2003tr}.)

The paper is organized as follows:
In the next section we give the relevant terms of the Lagrangian.
In section \ref{sec:3} we present the results of our calculation
in compact form using the formalism of \cite{K&T}.
Section \ref{sec:4} contains the formulae for various decay distributions.
We propose several T-odd asymmetries in section \ref{sec:5}. 
In section \ref{sec:6} we perform a numerical analysis of the
T-odd asymmetries proposed.
Finally, we summarize and conclude in section \ref{sec:7}.

\section{Lagrangian and couplings\label{sec:2}}

\noi
The terms of the Lagrangian necessary to calculate
the T-odd asymmetries and the  decay rates of
$\ti t_m \to \tilde{\chi}^0_kt$ and
$\tilde{\chi}^0_k \to \tilde{l}^\mp_nl^\pm_1 \to l^\pm_1l^\mp_2 \tilde\chi^0_1$
in the presence of the CP phases are:
\begin{equation}
\mathscr{L}_{l\tilde{l}\tilde{\chi}^0} \,= 
 g\,\bar{l}\,(a_{nk}^{\ti l}\,\PR + b_{nk}^{\ti l}\,\PL)\,\nt_k\,\ti l_n
+ {\rm h.c.}\ ,
\end{equation}
\begin{equation}
\mathscr{L}_{t\tilde{t}\tilde{\chi}^0}
= g\,\bar{t}\,(a_{mk}^{\st}\,\PR + b_{mk}^{\st}\,\PL)\,\nt_k\,\ti t_m
+ {\rm h.c.}\ ,
\end{equation}
where $P_{L,R}=\frac{1}{2}(1\mp \gamma_5)$,
$g$ is the $\mathrm{SU(2)}_L$ gauge coupling constant
and the couplings are defined as
\begin{equation}
a_{nk}^{\ti l}=({\mathcal R}^{\ti l}_{n1})^* f_{Lk}^l
+({\mathcal R}^{\ti l}_{n2})^* h_{Rk}^l~,\qquad
b_{nk}^{\ti l}=({\mathcal R}^{\ti l}_{n1})^* h_{Lk}^l
+({\mathcal R}^{\ti l}_{n2})^* f_{Rk}^l~,
\end{equation}
with ${\mathcal R}^{\ti l}_{nj}$ being the scalar lepton mixing
matrix and
\begin{eqnarray}
f_{Lk}^l & = &
{1\over \rzw}\Bigl( N_{k2}+\tan\tW N_{k1}\Bigr)\ ,
\hspace{-1cm} \label{eq:fLkl}\nonumber \\
f_{Rk}^l & = &
- {\rzw}\tan\tW N^{\ast}_{k1}\ ,\nonumber \\
h_{Rk}^l & = &(h_{Lk}^l)^*=Y_l N_{k3}~,
\end{eqnarray}
and
\be{eq:abneut}
a_{mk}^{\st} = \sum^2_{n=1}\, (\Rst_{mn})^{\ast}\,{\cal A}_{kn}^t, \qquad
b_{mk}^{\st} = \sum^2_{n=1}\, (\Rst_{mn})^{\ast}\,{\cal B}_{kn}^{t}\ .
\ee
Here ${\mathcal R}^{\ti t}_{mn}$ is the mixing matrix of the top squarks and 
\begin{equation}\label{eq:abneut1}
  {\cal A}_k^t = {f_{Lk}^t \choose h_{Rk}^t}, \qquad
  {\cal B}_k^t = {h_{Lk}^t \choose f_{Rk}^t},
\end{equation}
with
\baq{eq:htLk}
  f_{Lk}^t & = &
- {1\over \rzw}\Bigl( N_{k2}+{1\over 3}\tan\tW N_{k1}\Bigr)\ ,
    \hspace{-1cm} \label{eq:fLkt}\nonumber \\
  f_{Rk}^t & = &
    {2\rzw\over3}\tan\tW N^{\ast}_{k1}\ , \nonumber \\
  h_{Lk}^t & = & (h_{Rk}^t)^{\ast} = - Y_t N^{\ast}_{k4}\ .
\eaq
The unitary 4$\times$4 neutralino mixing matrix $N$ is defined in
Appendix A, Eq.~(\ref{eq:mixN}), $Y_t = m_t/(\sqrt{2}\:m_W\sin\b)$
and $Y_l = m_l/(\sqrt{2}\:m_W\cos\b)$.
The top squark mixing matrix $\Rst$ is given
in Appendix B, the scalar lepton mixing matrix ${\mathcal R}^{\ti l}$ can be
found, for instance, in \cite{Bartl:2002bh}.

\section{Formalism\label{sec:3}}

According to the formalism of \cite{K&T}
the differential decay rate of (\ref{st})-(\ref{chik}), 
when spin-spin correlations are taken into account, is:
\begin{eqnarray}
d\Gamma = d\Gamma  (\widetilde t_m \to t \tilde\chi^0_k)\,\frac{E_t}{m_t\Gamma_t}\,
d\Gamma (t\to ...)\,\frac{E_{\chi_k}}{m_{\chi_k} \Gamma_{\chi_k}}
\,d\Gamma (\tilde\chi^0_k\to ...),\label{dGamma}
\end{eqnarray}
where the factors $E_{\chi_k}/m_{\chi_k} \Gamma_{\chi_k}$
and $E_t/m_t\Gamma_t $ stem from
the used narrow width approximation for $t$ and $\tilde\chi^0_k$,
$\Gamma_t$ and $\Gamma_{\chi_k}$ are the total widths of the particles
and $m_{\chi_k}$ and $m_t$ are their masses.
We have
\begin{equation}
d\Gamma  (\widetilde t_m \to t  \tilde\chi^0_k)
=\frac4{2 m_{\ti t_m}}\vert A\vert^2\,d\Phi_{\tilde t}\,,\label{gammast}
\end{equation}
where
\begin{equation}
d\Phi_{\tilde t}=\frac{(2\pi)^4}{(2\pi)^6}\,\delta 
(p_{\ti t_m} - p_t-p_{\chi_k})\,
\frac{d\mathbf{p}_t}{2E_t}
\frac{d\mathbf{p}_{\chi_k}}{2 E_{\chi_k}}
\end{equation}
is the differential decay rate of the top squark $\tilde t_m$ into a top
quark with polarization 4-vector $\xi_t^\alpha$,
and a neutralino $\tilde \chi^0_k$ with polarization 4-vector $\xi_{\chi_k}^\alpha$.
For the matrix element $A$ we have:
\begin{eqnarray}
A&=&g\bar u(p_t) (b_{mk}^{\tilde t}P_L + a_{mk}^{\tilde t}P_R)v(p_{\chi_k}).
\end{eqnarray}
In evaluating $\vert A\vert^2$ we  use the spin density matrices of $t$
and $\tilde\chi^0_k$:
\begin{equation}
\rho (p_t) = \Lambda (p_t)\,\frac{1+\gamma_5\slashed{\xi}_t}{2},\qquad
\rho (-p_{\chi_k}) = - \Lambda (-p_{\chi_k})\,\frac{1+
\gamma_5\slashed{\xi}_{\chi_k}}{2},
\end{equation}
with
\begin{equation}
\Lambda (p_t) = \slashed{p}_t +m_t,\qquad \Lambda (p_{\chi_k}) =
\slashed{p}_{\chi_k}+m_{\chi_k},
\end{equation}
where $p_t$ and $p_{\chi_k}$ are the momentum 4-vectors
of the top quark and the neutralino $\tilde{\chi}^0_k$.
We have:
\begin{eqnarray}
\vert A\vert^2&=& \frac{g^2}{2}
\left\{ (\vert a_{mk}^{\tilde t}\vert^2 + \vert b_{mk}^{\tilde t}\vert^2)\,
[(p_{\chi_k} p_t) +
m_{\chi_k} m_t \, (\xi_{\chi_k}\xi_t)]\right.\nonumber\\
&&\qquad -(\vert a_{mk}^{\tilde t}\vert^2 - \vert b_{mk}^{\tilde t}\vert^2)\,
[m_t (p_{\chi_k}\xi_t)+m_{\chi_k}\,(\xi_{\chi_k} p_t)]\nonumber\\
&&\qquad -\,2\,{\Re e}(a_{mk}^{\tilde t*}b_{mk}^{\tilde t})\,
[m_{\chi_k} m_t -(p_{\chi_k}\xi_t)(\xi_{\chi_k} p_t)+
(p_{\chi_k} p_t)(\xi_{\chi_k} \xi_t)] \nonumber\\
&&\qquad +\left.2\, {\Im m}(a_{mk}^{\tilde t*}b_{mk}^{\tilde t})\,
\varepsilon (p_{\chi_k}\xi_{\chi_k}\xi_t p_t)\right\},
\label{A2}
\end{eqnarray}
where $\varepsilon^{0123}=1$.
The polarization 4-vectors $\xi_t^\alpha$ and $\xi_{\chi_k}^\alpha$
are determined through the decay processes
of the top quark and the neutralino.
$d\Gamma (t \to ...)$ and $d\Gamma (\tilde\chi^0_k\to ...)$ are the
differential decay rates of the
unpolarized top and unpolarized neutralino.

Next we shall consider the decays of $\tilde\chi^0_k$ and $t$. According
to the chosen decay mode
of the top quark, Eq. (\ref{t}), we have to distinguish two cases.
We consider them separately.

\subsection{Decay rates for $\tilde\chi^0_k\to l_1^\pm\tilde l_n^\mp$ }

For the width of the neutralino decay into a lepton $l^+_1$ and 
a scalar lepton $\tilde l_n$ we write
\begin{equation}
d\Gamma (\tilde\chi^0_k \to l_1^+\tilde l_n^-)=\frac{1}{2\cdot2 E_{\chi_k}}\,
{\rm Tr} (\bar B(-\Lambda (-p_{\chi_k}) )B)\,d\Phi_{\chi_k},\label{gammachi}
\end{equation}
with
\begin{equation}
d\Phi_{\chi_k}=\frac{(2\pi)^4}{(2\pi)^6}\,\delta
(p_{\chi_k} -{l_1^+} - p_{\ti l})\,
\frac{d\mathbf{p}_{\ti l}}{2 E_{\ti l}}
\frac{d\mathbf{l}^+_1}{2 E_{+}}.
\end{equation}
Here  $B$ is defined through the decay matrix element:
\begin{eqnarray}
\bar v_\sigma(p_{\chi_k})B^\sigma &=&g\,
\bar v(p_{\chi_k})(b_{nk}^{\tilde l*} P_R + a_{nk}^{\tilde l*} P_L)v({l_1^+}),
\end{eqnarray}
$p_{\chi_k}$ and ${l_1^+}$ are the momentum 4-vectors of
the neutralino and the lepton, $E_{\ti l}$ and $E_+$ are the energies
of $\ti l_n$ and $l_1^+$.
For the distribution of the decay products we obtain:
\begin{eqnarray}
d\Gamma (\ti\chi^0_k \to l_1^+\tilde l_n^-)=
\frac{g^2}{2\,E_{\chi_k}}(\vert a_{nk}^{\ti l}\vert^2+\vert 
b_{nk}^{\ti l}\vert^2)
(p_{\chi_k} {l_1^+})\,d\Phi_{\chi_k}.\label{gammak}
\end{eqnarray}
For the polarization vector $\xi^\alpha_{\chi_k}$ of
the neutralino $\tilde\chi^0_k$,
determined through the $\tilde\chi^0_k$-decay, we have:
\be{eq:xislepton}
\xi^\alpha_{\chi_k} =\left(g^{\alpha\beta} -
\frac{p_{\chi_k}^{\,\alpha} p_{\chi_k}^{\,\beta}}{m_{\chi_k}^2}\right)
\frac{{\rm Tr} (\bar B(-\Lambda (-p_{\chi_k}))\gamma_5 \gamma_\beta  B)}
{{\rm Tr} (\bar B(-\Lambda (-p_{\chi_k}) )B)}
 =\alpha_+\,\frac{m_{\chi_k}}{(p_{\chi_k}{l_1^+})}\,Q_+^\alpha ,
\ee
with
\be{eq:defQplus}
Q_+^\alpha =\left((l_1^+)^\alpha -\frac{(p_{\chi_k}{l_1^+})}
{m_{\chi_k}^2} p_{\chi_k}^\alpha \right),\qquad
 \alpha_+ = \frac{\vert b_{nk}^{\ti l}\vert^2-\vert a_{nk}^{\ti l}\vert^2}
{\vert b_{nk}^{\ti l}\vert^2+\vert a_{nk}^{\ti l}\vert^2}.
\ee
Respecting the condition ($\xi_{\chi_k}p_{\chi_k}) =0$, the vector $Q_+^\alpha$
is orthogonal to the
momentum 4-vector of $\tilde\chi^0_k$. This is the only orthogonal 4-vector 
composed of the available momenta $p_{\chi_k}$ and ${l_1^+}$.
As it can be seen from \rf{eq:xislepton} and \rf{eq:defQplus}, $\xi_{\chi_k}$
is in the $\tilde\chi^0_k$-decay plane. Further we shall assume that 
$\tilde l_n$ is produced on mass-shell,
$p_{\ti l}^2 =m_{\ti l}^2$, then $(p_{\chi_k}{l_1^+}) =
(m_{\chi_k}^2-m_{\ti l}^2)/2$, where we neglect the lepton mass in the
kinematics, i.e. $m_l=0$. The prefactor
$\alpha_+$ determines the sensitivity to the polarization of $\ti\chi^0_k$.

Note that the polarization vector of $\tilde\chi^0_k$, Eq.~\rf{eq:xislepton},
does not change if we take the subsequent decay $\ti l^-_n\to \ti\chi^0_1 l_2^-$
into account. Note further,  that
the polarization vector of the
C-conjugated decay $\tilde\chi^0_k \to l_1^-\tilde l_n^+$
changes sign compared to $\xi_{\chi_k}$ in Eq.~\rf{eq:xislepton}.

\subsection{Decay rate for $t\to b W^+$}

When the top quark decays according to $t\to bW^+$ we have:
\begin{equation}
d\Gamma (t \to bW^+)=\frac{1}{2 \cdot 2E_t}\, {\rm Tr} (\bar C_b \Lambda (p_t) C_b)
\,d\Phi_t^b\,,\label{gammatb}
\end{equation}
with
\begin{equation}
d\Phi_t^b=\frac{(2\pi)^4}{(2\pi)^6}\,\delta ( p_t - p_b- p_W )\,
\frac{d\mathbf{ p}_b}{2 E_b}\,
\frac{d\mathbf{ p}_W}{2 E_W},
\end{equation}
where $C_b$ is defined by the decay matrix element as follows:
\begin{eqnarray}
\bar C_b^\sigma u_\sigma (p_t)=\frac{g}{\sqrt 2}\,
\bar u(p_b )\gamma_\alpha P_L u(p_t)\,{\epsilon^\alpha}^*(p_W),\label{Cb}
\end{eqnarray}
where $p_b$, $p_t$ and $p_W$ are the momentum 4-vectors
of the bottom quark, the top quark and the $W$ boson.
Then for the distribution of the decay products we obtain:
\begin{eqnarray}
d\Gamma (t \to b W^+)= \frac{g^2}{8E_t}\, \frac{(m_t^2-m_W^2) \,
(2m_W^2+m_t^2)}{m_W^2}\,\,d\Phi_t^b.\label{gammab}
\end{eqnarray}
We denote by $\xi_b$ the polarization 4-vector of the top quark, 
determined by the decay $t\to bW$.
Its expression is  given by the formula:
\be{eq:xib}
\xi_b^\alpha =\left(g^{\alpha\beta} -\frac{p_t^\alpha p_t^\beta}{m_t^2}\right)
\frac{{\rm Tr} (\bar C_b\Lambda (p_t)\gamma_5 \gamma_\beta  C_b)}
{{\rm Tr} (\bar C_b \Lambda (p_t) C_b)}.
\ee
From (\ref{Cb}) and \rf{eq:xib} we obtain the polarization vector:
\begin{equation}
\xi_b^\alpha =\alpha_b\,\frac{m_t}{(p_tp_b)}\,Q_b^\alpha,\label{xitopb}
\end{equation}
with
\begin{equation}
\alpha_b= \frac{m_t^2-2m_W^2}{m_t^2+ 2m_W^2},\qquad
(p_tp_b) = \frac{m_t^2-m_W^2}{2},\qquad
Q_b^\alpha =\left(p_b^\alpha -\frac{(p_tp_b)}{m_t^2}\,p_t^\alpha \right),
\label{eq:alphab}
\end{equation}
where in the kinematics we have set $m_b=0$.
Here $Q_b^\alpha$ is the 4-vector  orthogonal to $p_t^\alpha$
and $\alpha_b$ determines the sensitivity to the polarization
of the top quark.

\subsection{Decay rates for $t\to b l \nu$ and for $t\to b c s$}

We consider here the decay $t\to b l \nu$.
For the decay $t\to b c s$ one has to make the replacements
$p_{\nu}\to p_s, p_l\to p_c$ in the equations below.
For the inclusion of QCD corrections to the decay
of a polarized top quark we refer to \cite{Czarnecki:1990pe}.
When the top quark decays according to $t\to bl\nu$ we have:
\begin{equation}
d\Gamma (t \to bl\nu)=\frac{1}{2\cdot 2E_t}\, {\rm Tr} (\bar C_l \Lambda (p_t) C_l)
\,d\Phi_t^l\,,\label{gammatl}
\end{equation}
with
\be{eq:psPhil}
d\Phi_t^l=\frac{(2\pi)^4}{(2\pi)^9}\,\delta (p_t - p_b- p_{l} - p_\nu )\,
\frac{d\mathbf{ p}_b}{2 E_b}\,
\frac{d\mathbf{ p}_{l}}{2 E_{l}}\,\frac{d\mathbf{ p}_\nu}{2 E_\nu },
\ee
where $C_l$ is defined through the decay martix element as follows:
\begin{eqnarray}
\bar C_l^\sigma u_\sigma (p_t)=-i\,(\frac{g}{\sqrt 2})^2\bar u (p_\nu )
\gamma_\alpha P_L v(p_{l})\,
\frac{g^{\alpha\beta} -p_W^\alpha p_W^\beta/m_W^2}{D_W}\,
\bar u(p_b )\gamma_\beta P_L u(p_t),\label{Cl}
\end{eqnarray}
with
\begin{eqnarray}
D_W =(p_W^2-m_W^2) + im_W\Gamma_W,\qquad p_W^{\,\alpha} = p_t^{\,\alpha}
-p_b^{\,\alpha}.
\end{eqnarray}
Then we obtain:
\begin{eqnarray}
d\Gamma (t \to bl\nu)=\frac{g^4}{2E_t |D_W|^2} (p_tp_l)
\,(m_t^2-2(p_tp_l))\,d\Phi_t^l.
\end{eqnarray}
From (\ref{Cl}), for the polarization vector of the top, that we denote
by $\xi_l$, we have:
\be{eq:xitopl}
\xi_l^\alpha =\left(g^{\alpha\beta} -\frac{p_t^\alpha p_t^\beta }{m_t^2}\right)
\frac{{\rm Tr} (\bar C_l\Lambda (p_t)\gamma_5 \gamma_\beta  C_l)}
{{\rm Tr} (\bar C_l \Lambda (p_t) C_l)}
=\alpha_l\,\frac{m_t}{(p_tp_l)}\,Q_l^\alpha ,
\ee
with
\be{eq:alphap}
Q_l^\alpha =\left(p_l^\alpha -\frac{(p_tp_l)}{m_t^2}\,p_t^\alpha \right),\qquad
\alpha_l=-1,
\ee
$Q_l^\alpha$ is orthogonal to $p_t^\alpha$ and lays in the
top quark decay plane (in the rest frame of the top quark).
In general,
with the available vectors in the decay, $p_t$, $p_b$ and $p_l$,
one can form three independent
combinations orthogonal to $p_t^\alpha$: two in the decay plane,
$Q_b^\alpha$ and $Q_l^\alpha$, and one transverse to it,
$\varepsilon (\alpha p_tp_bp_l)$. As CP invariance holds
in the top quark decay,
there is no contribution to the transverse component, and
 because of the V-A structure of  the interaction, there is
no contribution to  $Q_b^\alpha$ either.

Inserting (\ref{gammast}), (\ref{gammachi}) and (\ref{gammatb})
or (\ref{gammatl}) into (\ref{dGamma})
we obtain $d\Gamma$ in terms of the polarization vectors:
\begin{eqnarray}
d\Gamma^{b,l} &=& \frac{1}{2 m_{\ti t_m}}\frac{1}{2m_t\Gamma_t}
\frac{1}{2 m_{\chi_k} \Gamma_{\chi_k}}
\vert A\vert^2\,{\rm Tr} (\bar B(-\Lambda (-p_{\chi_k}) )B)\,
{\rm Tr} (\bar C_{b,l} \Lambda (p_t) C_{b,l})
\nonumber\\
&&{}\times \frac{1}{2 m_{\ti l} \Gamma_{\ti l}}g^2
(|a_{n1}^{\ti l}|^2+|b_{n1}^{\ti l}|^2)
(m_{\ti l}^2-m_{\chi_1}^2) 
\,d\Phi^{b,l},\label{dGammaxi}
\end{eqnarray}
where we have used the narrow width approximation for the scalar
lepton propagator.
$m_{\ti l}$ and $\Gamma_{\ti l}$ is the mass and the total
decay width of $\ti l$,
$\vert A\vert^2$ is given by (\ref{A2}) and $d\Phi^{b,l}$
denotes the phase space for the two different decay modes of the top quark:
\be{eq:phasespace}
d\Phi^{b,l}=d\Phi_{\tilde t} \cdot d\Phi_t^{b,l} \cdot d\Phi_{\chi_k}
\cdot d\Phi_{\ti l}.
\ee
In order to obtain the angular distributions of the ordinary particles
in (\ref{st}) - (\ref{chik}) we have to use the
explicit expressions for $\xi_{\chi_k}$ and $\xi_t$, and carry
the integration over the phase space of the supersymmetric particles.

\section{Decay distributions\label{sec:4}}

In this section we derive the analytical expressions for the decay
distributions of 
$\ti t_m\to t \ti\chi^0_k\to b W^+~ \ti\chi^0_1 
l^{\pm}_1 l^{\mp}_2$
and 
$\ti t_m\to t \ti\chi^0_k\to b l \nu \ti\chi^0_1 
l^{\pm}_1 l^{\mp}_2$.
We consider separately the two decays (\ref{t}) of the top quark.

\subsection{Decay distribution for 
$\ti t_m\to t \ti\chi^0_k\to b W^+~ \ti\chi^0_1 
l^{\pm}_1 l^{\mp}_2$}

We choose $\mathbf p_t$ in the direction of the $Z$-axis and
$\mathbf p_t$ and $\mathbf p_b$ determine the YZ-plane:
\baq{eq:kinematics}
&&p_{\ti t_m} = (m_{\ti t_m},\overrightarrow 0),\qquad
{\mathbf{p}}_t = \vert {\mathbf p}\vert\, (0,0,1),
\qquad {\mathbf{p}}_{\chi_k} =\vert {\mathbf p}\vert\, (0,0,-1),\nonumber\\
&&{}\!\!\!\!\!\!\!\!\!\!\!\!\!\!\!\!
{\mathbf p}_b = E_b \,(0,s_b, c_b),\quad 
{\mathbf l}^+_1
= E_+ (s_+c_{\phi_+},s_+s_{\phi_+},c_+),\quad 
{\mathbf l}^-_2
= E_- (s_-c_{\phi_-},s_-s_{\phi_-},c_-),\nonumber\\
\eaq
where we have used the brief notation
$c_b=\cos\theta_b$, $s_{\phi_+}=\sin\phi_+$, etc.
The ranges of the angles are
$0 \leq\theta_b,\theta_+,\theta_-\leq \pi$; 
$0 \leq \phi_+,\phi_-\leq 2 \pi$.
Then we can carry out part of the phase space integration.
Using \rf{eq:kinematics}, \rf{eq:phasespace} is given by
\begin{eqnarray}
d\Phi^b =\frac{\vert \mathbf p\vert\, (m_t^2-m_W^2)
(m_{\chi_k}^2 -m_{\ti l}^2)}
{2\,m_{\ti t_m}\,8^2(2\pi )^4\,E_t^2\, E_{\chi_k}^2}
\,\frac{dc_b\,dc_+\,d\phi_+}{(1 -\beta_tc_b)^2(1+\beta_{\chi_k}c_+)^2}
\cdot d\Phi_{\ti l},
\end{eqnarray}
where
\be{eq:psslep}
d\Phi_{\ti l}=\frac{1}{8(2\pi)^2}
\frac{m_{\ti l}^2-m_{\chi_1}^2}{E_{\ti l}^2
(1-\beta_{\ti l}c_{\ti l l^-})^2}~ 
d\Omega_-,
\ee
and
\begin{eqnarray}
\beta_t = \frac{\vert \mathbf p\vert}{E_t},\qquad  \beta_{\chi_k}
= \frac{\vert \mathbf p\vert}{E_{\chi_k}},\qquad
\beta_{\ti l} =\frac{|{\bf p}_{\ti l}|}{E_{\ti l}},\qquad
\vert {\mathbf p}\vert\,= \frac{\lambda^{1/2} (m_{\ti t_m}^2, m_t^2,
m_{\chi_k}^2)}{2\, m_{\ti t_m}},\nonumber
\end{eqnarray}
\vspace{-0.8cm}
\begin{eqnarray}
E_t = \sqrt{\vert {\mathbf p}\vert^2+m_t^2} =\frac{m_{\ti t_m}^2
+m_t^2 -m_{\chi_k}^2}{2 m_{\ti t_m}},\nonumber
\end{eqnarray}
\vspace{-0.8cm}
\begin{eqnarray}
E_{\chi_k} = \sqrt{\vert {\mathbf p}\vert^2+m_{\chi_k}^2} =\frac{m_{\ti t_m}^2
-m_t^2 +m_{\chi_k}^2}{2  m_{\ti t_m}},\hspace{0.3cm}
c_{\ti l l^-}=(\hat {\mathbf p}_{\ti l}\cdot \hat {\mathbf p}_{l_2^-}),
\end{eqnarray}
where $\lambda(x,y,z)=x^2+y^2+z^2-2(x y+x z+ yz)$.
Then from (\ref{dGammaxi}),  using the explicit expressions
for the polarization vectors \rf{eq:xislepton} and (\ref{xitopb}) and
the decay distributions (\ref{gammak}) and (\ref{gammab}),
for the angular  distributions of the $b$-quark and the leptons $l_1^+$ and
$l_2^-$, we obtain:
\baq{dGammab}
\frac{d^5\Gamma_b}{dc_b\,d\Omega_+\,d\Omega_-}&=&N_{\ti l}
\frac{1}{E^2_{\ti l}(1-\beta_{\ti l}\, c_{\ti l l^-})^2}
N_b\,\frac{1}{(1-\beta_tc_b)^2(1+\beta_{\chi_k}c_+)^2}\nonumber\\
&&{}\!\!\!\!\!\!\!\!\!\!\!\!\!\!\!\!\!\!
\times\left\{ (\vert a_{mk}^{\tilde t}
\vert^2 + \vert b_{mk}^{\tilde t}\vert^2)\,
\left [(p_{\chi_k} p_t)\,+ \alpha_b\,\alpha_+\,\frac{m_t^2}{(p_tp_b)}
\frac{m_{\chi_k}^2}{(p_{\chi_k}l^+_1)}
\, (Q_+Q_b)\right]\right.\nonumber\\
&&{}\!\!\!\!\!\!\!\!\!\!\!\!\!\!\!\!\!\!
-(\vert a_{mk}^{\tilde t}\vert^2 - \vert b_{mk}^{\tilde t} \vert^2)
\left[\alpha_b\,\frac{m_t^2}{(p_tp_b)}\,(Q_b p_{\chi_k})\,+ \,\alpha_+\,
\frac{m_{\chi_k}^2}{(p_{\chi_k}l^+_1)}
\,(Q_+ p_t)\,\right]\nonumber\\
&&\!\!\!\!\!\!\!\!\!\!\!\!\!\!\!\!\!\!
-\,2\,{\Re e}(a_{mk}^{\tilde t*}b_{mk}^{\tilde t})\,m_{\chi_k} m_t
\nonumber\\
&&{}\!\!\!\!\!\!\!\!\!\!\!\!\!\!\!\!\!\!
\times\left[1\,-\frac{\alpha_b}{(p_tp_b)}\frac{\alpha_+}{(p_{\chi_k}l_1^+)}
\left[(p_{\chi_k} Q_b)\,(Q_+ p_t)-
(p_{\chi_k} p_t)\,(Q_+Q_b)\right]\right]
\nonumber\\
&&{}\!\!\!\!\!\!\!\!\!\!\!\!\!\!\!\!\!\!
\left.+ 2\, {\Im m}(a_{mk}^{\tilde t*}b_{mk}^{\tilde t})\,
\alpha_b\,\,\alpha_+\,\frac{m_t}{(p_tp_b)}\,
\frac{m_{\chi_k}}{(p_{\chi_k} l_1^+)}\,m_{\ti t_m}
\,\left(\mathbf{l}_1^+\mathbf{ p}_b \mathbf{ p}_t\right)\right\},
\eaq
where
\begin{eqnarray}
N_{\ti l}&=&\frac{1}{m_{\ti l}\Gamma_{\ti l}}\frac{\alpha_w}{8}\frac{1}{4\pi}
(|a_{n1}^{\ti l}|^2+|b_{n1}^{\ti l}|^2)(m^2_{\ti l}-m^2_{\chi_1})^2\,,\nonumber\\
N_b&=&\left(\frac{\alpha_w}{8}\right)^3\,\frac{(2m_W^2+m_t^2)(m_t^2-m_W^2)^2
(m_{\chi_k}^2 -m_{\ti l}^2)^2\,(|a_{nk}^{\tilde l}|^2+|b_{nk}^{\tilde l}|^2)
\,\vert {\mathbf p}\vert}
{2\pi\, m_{\ti t_m}^2\,m_W^2\,m_t\,\Gamma_t m_{\chi_k}\,\Gamma_{\chi_k}
\,E_t^2\,E_{\chi_k}^2},\nonumber\\
E_b&=& \frac{m_t^2-m_W^2}{2\,E_t\,(1-\beta_tc_b)},~ E_+=
\frac{m_{\chi_k}^2-m_{\ti l}^2}{2\,E_{\chi_k}\,(1+\beta_{\chi_k}c_+)},~
E_-=\frac{m^2_{\ti l}-m_{\chi_1}^2}{2E_{\ti l}
(1-\beta_{\ti l}c_{\ti l l^-})},\nonumber\\
\left(\mathbf{l}_1^+\mathbf{ p}_b \mathbf{ p}_t\right)&=& E_+E_b
\vert {\mathbf p}\vert s_bs_+c_{\phi_+}.
\end{eqnarray}

\subsection{Decay distribution of 
$\ti t_m\to t \ti\chi^0_k\to b l \nu \ti\chi^0_1 
l^{\pm}_1 l^{\mp}_2$}\label{sec:3.2}

The  angular distribution of the final $b$-quark and leptons $l$, $l_1^+$
and $l_2^-$ is obtained from the previous results if
we fix the coordinate system so that ${\mathbf p}_t$ and ${\mathbf p}_l$
determine the YZ-plane:
\begin{eqnarray}
&&
{\mathbf{p}}_t = \vert {\mathbf p}\vert\, (0,0,1),
\qquad {\mathbf{p}}_{\chi_k} =\vert {\mathbf p}\vert\, (0,0,-1),\qquad
{\mathbf p}_l = E_l \,(0,s_l, c_l),\nonumber\\
&&{\mathbf p}_b = E_b \,(s_bc_{\phi_b},s_bs_{\phi_b}, c_b),
\end{eqnarray}
where the ranges of the angles are
$0 \leq\theta_b,\theta_l\leq \pi$; $0 \leq \phi_b\leq 2 \pi$.
Then the dependence on the $b$-quark momentum is only in the phase space. 
We obtain:
\begin{eqnarray}
d\Phi^l &=& \frac{\vert \mathbf p\vert\, m_W^2\,
(m_t^2-m_W^2)(m_{\chi_k}^2 -m_{\ti l}^2)}
{2\,m_{\ti t_m}\,8^3(2\pi )^7\,E_t^2\,E_{\chi_k}^2}\,\nonumber\\
&&\times\,\frac{dc_l\,d\Omega_b\,d\Omega_+}{(1 -\beta_tc_b)^2
(1+\beta_{\chi_k}c_+)^2
[E_t(1-\beta_tc_l)-E_b(1-c_{bl})]^2}\cdot \frac{ds_W}{2 \pi} 
\cdot d\Phi_{\ti l},\nonumber\\
\end{eqnarray}
where $s_W=p^2_W$.
We obtain the angular distribution by a replacement of the phase space
$d\Phi^b \rightarrow d\Phi^l$ and the following replacements
in the curly brackets of \rf{dGammab}:
$\alpha_b\rightarrow\alpha_l$, $Q_b \rightarrow Q_l$ and
$p_b\rightarrow p_l$.
The angular decay rate distribution of $l_1^+$, $l_2^-$, $l$ and $b$ is:
\baq{dGammal}
 \frac{d^7\Gamma_l}{dc_l\,d\Omega_b\,d\Omega_+\,d\Omega_-}&=&
N_{\ti l}
\frac{1}{E^2_{\ti l} (1-\beta_{\ti l}\, c_{\ti l l^-})^2}
\nonumber\\
&&{}\!\!\!\!\!\!\!\!\!\!\!\!\!\!\!\!\!\!
\times
N_l\,
\frac{(p_tp_l)(m_t^2 -2(p_tp_l))}
{(1-\beta_tc_b)^2(1+\beta_{\chi_k}c_+)^2[E_t(1-\beta_tc_l) -
E_b (1-c_{bl})]^2}\nonumber\\
&&{}\!\!\!\!\!\!\!\!\!\!\!\!\!\!\!\!\!\!
\times\left\{(\vert a_{mk}^{\tilde t}\vert^2 + \vert
b_{mk}^{\tilde t}\vert^2)\,
\left[(p_{\chi_k} p_t)\,+ \alpha_l\,\alpha_+\,\frac{m_t^2}{(p_tp_l)}
\frac{m_{\chi_k}^2}{(p_{\chi_k}l_1^+)}
 \, (Q_+Q_l)\right]\right.
\nonumber\\
&&{}\!\!\!\!\!\!\!\!\!\!\!\!\!\!\!\!\!\!
-(\vert a_{mk}^{\tilde t}\vert^2 - \vert b_{mk}^{\tilde t}
\vert^2)\,
 \left[\alpha_l\,\frac{m_t^2}{(p_tp_l)}\,(Q_l p_{\chi_k})\,+ \,
\alpha_+\,\frac{m_{\chi_k}^2}{(p_{\chi_k}l_1^+)}\,(Q_+ p_t)\,\right]
\nonumber\\
&&{}\!\!\!\!\!\!\!\!\!\!\!\!\!\!\!\!\!\!
-\,2\,{\Re e}(a_{mk}^{\tilde t*}b_{mk}^{\tilde t})\,
m_{\chi_k} m_t \,
\nonumber\\
&&{}\!\!\!\!\!\!\!\!\!\!\!\!\!\!\!\!\!\!
\times\left[1\,-\frac{\alpha_l}{(p_tp_l)}\frac{\alpha_+}{(p_{\chi_k}l^+_1)}
\left[(p_{\chi_k}Q_l)\,(Q_+ p_t)- (p_{\chi_k} p_t)\,(Q_+Q_l)\right]\right]
\nonumber\\
&&{}\!\!\!\!\!\!\!\!\!\!\!\!\!\!\!\!\!\!
\left.+ 2\,{\Im m}(a_{mk}^{\tilde t*}b_{mk}^{\tilde t})\,
\alpha_l\,\,\alpha_+\,\frac{m_t}{(p_tp_l)}\,
\frac{m_{\chi_k}}{(p_{\chi_k}l^+_1)}\,m_{\ti t_m}
\,\left(\mathbf{l}_1^+\mathbf{ p}_l \mathbf{ p}_t\right)\right\},
\eaq
where
\begin{eqnarray}
N_l&=& \left(\frac{\alpha_w}{8}\right)^4\,\frac{m_W^2(m_t^2-m_W^2)\,
\vert {\mathbf p}\vert\,
(m_{\chi_k}^2 -m_{\ti l}^2)^2\,(|a_{nk}^{\tilde l}|^2+|b_{nk}^{\tilde l}|^2)}
{\pi^2\,m_{\ti t_m}^2\,m_t\,\Gamma_t m_{\chi_k}\,\Gamma_{\chi_k}\,
m_W\,\Gamma_W\,E_t^2\,E_{\chi_k}^2},\\
E_l&=&\frac{m_W^2}{2\left[E_t(1-\beta_tc_l)-E_b(1-c_{bl})\right]},\qquad
c_{b l}=(\hat {\mathbf p}_b\cdot \hat {\mathbf p}_l),\\
\left(\mathbf{l}_1^+\mathbf{ p}_l \mathbf{ p}_t\right)&=& E_+E_l
\vert {\mathbf p}\vert s_ls_+c_{\phi_+}.
\end{eqnarray}
The distribution of $l_1^-$ and $l_2^+$ from the C-conjugate decay
$\ti\chi^0_k \to l_1^-\ti l_n^+\to l_1^-l_2^+\ti\chi^0_1$ is obtained from
\rf{dGammab} and \rf{dGammal} by the replacements
$l_1^+ \rightarrow l_1^-$, $l_2^- \rightarrow l_2^+$ and
$\alpha_+ \rightarrow -\alpha_+$.

As can be seen from the angular distributions,
Eqs.~\rf{dGammab} and \rf{dGammal},
the prefactor of the triple product correlations
(last term in Eqs.~\rf{dGammab} and \rf{dGammal})
is ${\Im m}(a_{mk}^{\tilde t*}b_{mk}^{\tilde t})$ and consequently
the T-odd asymmetries (to be defined in the next section)
are propotional to this prefactor.
Therefore, in order to study the dependence of the T-odd asymmetries
on the MSSM parameters, it is useful to give the explicit expression
for ${\Im m}(a_{mk}^{\tilde t*}b_{mk}^{\tilde t})$ for $m=1$ using
Eqs.~\rf{eq:abneut}-\rf{eq:htLk}:
\baq{eq:expand}
{\Im m}(a_{1k}^{\tilde t*}b_{1k}^{\tilde t})&=&
-\cos^2\theta_{\ti t} Y_t {\Im m}({f^t_{Lk}}^* N_{k4}^*)
-\sin^2\theta_{\ti t} \frac{2 \sqrt{2}}{3} Y_t \tan\theta_W
{\Im m}(N_{k1}^* N_{k4}^*)
\nonumber\\
&&{}\!\!\!\!\!\!\!\!\!\!\!\!\!\!\!\!
+\cos\theta_{\ti t}\sin\theta_{\ti t}
\left(\frac{2 \sqrt{2}}{3}\tan\theta_W
{\Im m}({f^t_{Lk}}^* N_{k1}^* e^{i\phi_{\ti t}})
+Y^2_t {\Im m}(N_{k4}^* N_{k4}^* e^{-i\phi_{\ti t}})
\right).\nonumber\\
\eaq
We can see from \rf{eq:expand} that if CP violation
is solely due to $\phi_{A_t}\neq 0$, the T-odd asymmetries
are proportional to $\sin 2\theta_{\ti t} \sin\phi_{\ti t}$,
which can be naturally large because of the
large top squark mixing (see Eqs.~\rf{eq:stopmass} and (\ref{eq:mlr})).
Moreover, one can see from \rf{eq:expand} that
the term $\propto \sin 2\theta_{\ti t}$ can be sizable
also in a higgsino-like scenario ($|\mu|< M_2$)
because of the large top Yukawa coupling.

\section{T-odd asymmetries \label{sec:5}}

We shall distinguish three classes of asymmetries
according to the lepton momentum (stemming from the decay chain
$\ti\chi^0_k\to \ti l^{\mp}_n l_1^{\pm} \to l_1^{\pm}
l_2^{\mp}\ti\chi^0_1 $) involved in the triple product:
($i$) when the momentum vector
of lepton $l^{\pm}_1$ from the decay $\ti\chi^0_k\to \ti l^{\mp}_n
l_1^{\pm}$ enters; ($ii$) when the lepton momentum vector of lepton
$l^{\mp}_2$ from the decay $\ti l^{\mp}_n\to l_2^{\mp}\ti\chi^0_1$
enters; and ($iii$) when both momentum vectors of $l_1^{\pm}$ and
$l_2^{\mp}$ from the decay $\ti\chi^0_k\to l_1^{\pm}
l_2^{\mp}\ti\chi^0_1$ enter.\\
$\bullet$ The first class involves the asymmetries:
\be{eq:asym1}
A_1^{\pm}=\frac{ N\left[({\bf p}_b {\bf p}_t {\bf l}_1^{\pm}) > 0 \right]
 -N\left[({\bf p}_b{\bf p}_t {\bf l}_1^{\pm}) < 0 \right]}
 {N\left[({\bf p}_b{\bf p}_t {\bf l}_1^{\pm}) > 0 \right]
  +N \left[({\bf p}_b{\bf p}_t {\bf l}_1^{\pm}) < 0 \right]},
   \ee
\be{eq:asym2}
 A_2^{\pm}= \frac{ N\left[({\bf p}_{l}{\bf p}_t {\bf
l}_1^{\pm})  > 0 \right] -N\left[({\bf p}_{l}{\bf p}_t {\bf
l}_1^{\pm}) < 0 \right]} {N\left[({\bf p}_{l}{\bf p}_t {\bf
l}_1^{\pm}) > 0 \right] +N\left[({\bf p}_{l}{\bf p}_t {\bf
l}_1^{\pm}) < 0 \right]}, \ee
\be{eq:asym3}
A_3^{\pm}=
\frac{
N\left[({\bf p}_{l} {\bf p}_b {\bf l}_1^{\pm})  > 0 \right]
-N\left[({\bf p}_{l} {\bf p}_b {\bf l}_1^{\pm}) < 0 \right]}
{N\left[({\bf p}_{l} {\bf p}_b {\bf l}_1^{\pm}) > 0 \right]
+N\left[({\bf p}_{l} {\bf p}_b {\bf l}_1^{\pm}) < 0 \right]},
\ee
where ${\bf p}_{l}$ is the lepton momentum in the decay $t\to b l\nu$.\\
\noi
$\bullet$ In the second class of the asymmetries $l_1$ is
replaced by $l_2$:
\be{eq:asymp1}
A_1^{'\pm}=\frac{
N\left[({\bf p}_b {\bf p}_t {\bf l}_2^{\pm}) > 0 \right]
-N\left[({\bf p}_b{\bf p}_t {\bf l}_2^{\pm}) < 0 \right]}
{N\left[({\bf p}_b{\bf p}_t {\bf l}_2^{\pm}) > 0 \right]
+N\left[({\bf p}_b{\bf p}_t {\bf l}_2^{\pm}) < 0 \right]},
\ee
\be{eq:asymp2}
A_2^{'\pm}=
\frac{
N\left[({\bf p}_l{\bf p}_t {\bf l}_2^{\pm})  > 0 \right]
-N\left[({\bf p}_l{\bf p}_t {\bf l}_2^{\pm}) < 0 \right]}
{N\left[({\bf p}_l{\bf p}_t {\bf l}_2^{\pm}) > 0 \right]
+N\left[({\bf p}_l{\bf p}_t {\bf l}_2^{\pm}) < 0 \right]},
\ee
\be{eq:asymp3}
A_3^{'\pm}=
\frac{
N\left[({\bf p}_l {\bf p}_b {\bf l}_2^{\pm})  > 0 \right]
-N\left[({\bf p}_l {\bf p}_b {\bf l}_2^{\pm}) < 0 \right]}
{N\left[({\bf p}_l {\bf p}_b {\bf l}_2^{\pm}) > 0 \right]
+N\left[({\bf p}_l {\bf p}_b {\bf l}_2^{\pm}) < 0 \right]}.
\ee
$\bullet$
The third class of asymmetries is:
\be{eq:asym4}
A_4^{\pm}=
\frac{
N\left[({\bf p}_b{\bf l}_1^{\pm} {\bf l}_2^{\mp})  > 0 \right]
-N\left[({\bf p}_b{\bf l}_1^{\pm} {\bf l}_2^{\mp}) < 0 \right]}
{N\left[({\bf p}_b{\bf l}_1^{\pm} {\bf l}_2^{\mp}) > 0 \right]
+N\left[({\bf p}_b{\bf l}_1^{\pm} {\bf l}_2^{\mp}) < 0 \right]},
\ee
\be{eq:asym5}
A_5^{\pm}=
\frac{
N\left[({\bf p}_l{\bf l}_1^{\pm} {\bf l}_2^{\mp})  > 0 \right]
-N\left[({\bf p}_l{\bf l}_1^{\pm} {\bf l}_2^{\mp}) < 0 \right]}
{N\left[({\bf p}_l{\bf l}_1^{\pm} {\bf l}_2^{\mp}) > 0 \right]
+N\left[({\bf p}_l{\bf l}_1^{\pm} {\bf l}_2^{\mp}) < 0 \right]}.
\ee
Since the polarization vectors of $\tilde\chi^0_k$ for the two C-conjugate
decay modes of
(\ref{chik}) differ only by a sign (see Eq.~\rf{eq:xislepton})
the value of the asymmetries with upper indices  $+$ and $-$ are related by:
\be{eq:relation}
A_i^+=-A_i^- ~(i=1,\dots,5)\qquad  {\rm and}\qquad A_i^{'+}=-A_i^{'-} ~(i=1,2,3).
\ee
In order to measure all of the listed
asymmetries  it is necessary to distinguish the lepton
$l^{\pm}_1$, originating from the decay $\ti\chi^0_k\to \ti l^{\mp}_n
l_1^{\pm}$, and the lepton $l^{\mp}_2$ from the subsequent decay
$\ti l^\mp_n\to \ti\chi^0_1 l^{\mp}_2$. This can be accomplished by measuring 
the energies of the leptons and making use
of their different energy distributions, 
when the masses of the particles involved are known.
$l^{\pm}_1$ and $l^{\mp}_2$ can be 
be distinguished if their measured energies do
not lie in the overlapping region of their energy distributions.

 $\bullet$ We define the fourth class of asymmetries as follows:
\be{eq:asymchar1}
A_1=\frac{
N\left[({\bf p}_b {\bf p}_t {\bf l}^+) > 0 \right]
-N\left[({\bf p}_b{\bf p}_t {\bf l}^+) < 0 \right]}
{N\left[({\bf p}_b {\bf p}_t {\bf l}^+) > 0 \right]
+N\left[({\bf p}_b{\bf p}_t {\bf l}^+) < 0 \right]},
\ee
\be{eq:asymchar2}
A_2=\frac{
N\left[({\bf p}_l {\bf p}_t {\bf l}^+) > 0 \right]
-N\left[({\bf p}_l{\bf p}_t {\bf l}^+) < 0 \right]}
{N\left[({\bf p}_l {\bf p}_t {\bf l}^+) > 0 \right]
+N\left[({\bf p}_l{\bf p}_t {\bf l}^+) < 0 \right]},
\ee
\be{eq:asymchar3}
A_3=\frac{
N\left[({\bf p}_l {\bf p}_b {\bf l}^+) > 0 \right]
-N\left[({\bf p}_l{\bf p}_b {\bf l}^+) < 0 \right]}
{N\left[({\bf p}_l {\bf p}_b {\bf l}^+) > 0 \right]
+N\left[({\bf p}_l{\bf p}_b {\bf l}^+) < 0 \right]},
\ee
where ${\bf l}^+$ stands for the 
momentum 3-vector of either $l_1^+$ or $l_2^+$. Evidently we have
\be{eq:A}
A_i=\frac{A_i^+ +A_i^{'+}}{2}, \quad
~i=1,2,3.
\ee
A measurement of $A_i,~ i=1,2,3$, does not require to distinguish
between the leptons $l^\pm_1$ and $l^\mp_2$, it requires only a measurement
of their charges. Analogous asymmetries can be defined for $l^-$ as well.

It should also be noted that the asymmetries given in 
Eqs.~\rf{eq:asym1}-\rf{eq:asym3} do not depend on the mass of $\ti l_n$. 
The asymmetries above are written down for the leptonic decay 
$W^+\to l \nu$. For the hadronic decay $W^+\to c s$ the analogous
asymmetries are obtained by replacing ${\bf p}_l \to {\bf p}_c$.

\section{Numerical results \label{sec:6}}

All proposed T-odd asymmetries depend on 
${\Im m}(a_{mk}^{\tilde t*}b_{mk}^{\tilde t})$, Eq.~\rf{eq:expand},
and measure therefore the same combination of CP phases
in the MSSM, but they have different magnitude.
In this section we present numerical results for the
asymmetries $A^{\pm}_i, A'^{\pm}_i~ (i=1,2,3)$, Eqs.~\rf{eq:asym1}-\rf{eq:asymp3}, 
$A^{\pm}_4, A^{\pm}_5$, Eqs.~\rf{eq:asym4} and \rf{eq:asym5} and
$A_i~ (i=1,2,3)$, Eqs.~\rf{eq:asymchar1}-\rf{eq:asymchar3}.
In order not to vary too many parameters we 
fix $m_{\ti t_1}=400$ GeV, $m_{\ti t_2}=800$ GeV 
and $\tan\beta=10$\footnote{The prefactor 
${\Im m}(a_{mk}^{\tilde t*}b_{mk}^{\tilde t})$
of the asymmetries does not depend very much on the value
of $\tan\beta$ if $|A_t|\gg |\mu|/\tan\beta$.}.
We take $m_t=178$~GeV and we also use the GUT relation $|M_1|=5/3\tan^2\Theta_W M_2$.
We take $m_{\ti l_1}=130$~GeV, $m_{\ti l_2}=300$~GeV,
where we assume $\ti l_1\approx \ti l_R$ for $l=e,\mu$, which is
suggested in mSugra models.
In the scalar tau sector we take into account scalar tau mixing 
choosing $A_{\tau}=500$~GeV.
In our numerical study we take $|A_t|,
\phi_{A_t}$, $M_2$, $\phi_{M_1}$, $|\mu|,
\phi_{\mu}$ as input parameters. 
Note that for a given set of input parameters we obtain
two solutions for ($M_{\ti Q},M_{\ti U}$) corresonding to
the cases $M_{\st_{LL}}^2>M_{\st_{RR}}^2$ 
and $M_{\st_{LL}}^2<M_{\st_{RR}}^2$ in Eqs.~(\ref{eq:mll}) and
(\ref{eq:mrr}) which we will treat separately.
In the plots we impose the phenomenological constraints:
$m_{\ti\chi^{\pm}_1}>103$~GeV,$m_{\ti\chi^0_1}>50$~GeV
and $\ti\chi^0_1$ is the lightest supersymmetric particle (LSP).

\begin{figure}[t]
\setlength{\unitlength}{1mm}
\begin{center}
\begin{picture}(150,120)
\put(-53,-65){\mbox{\epsfig{figure=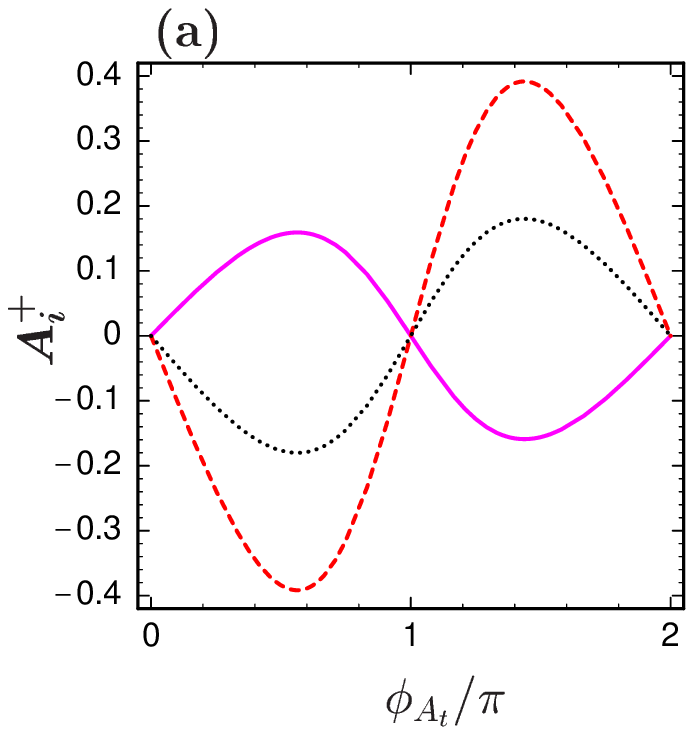,height=22.cm,width=19.4cm}}}
\put(27,-65){\mbox{\epsfig{figure=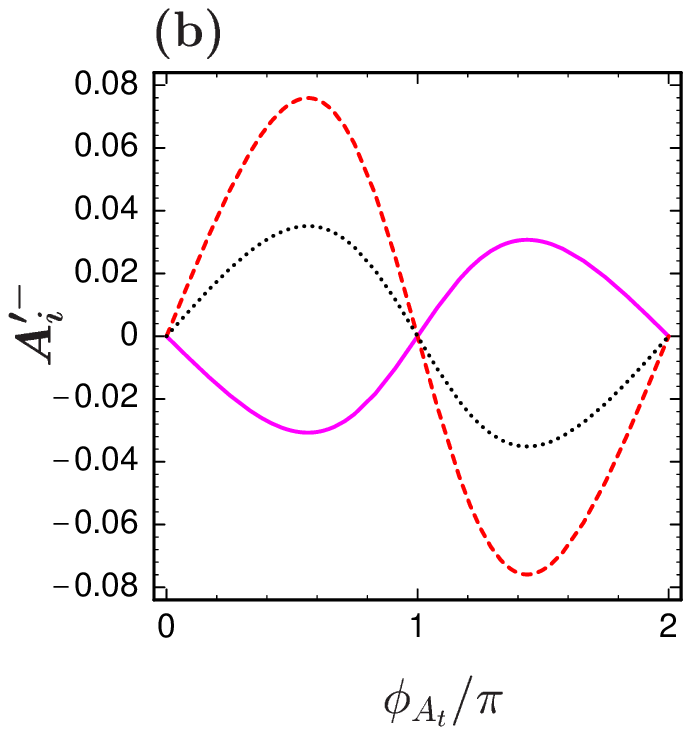,height=22.cm,width=19.4cm}}}
\put(-53,-125){\mbox{\epsfig{figure=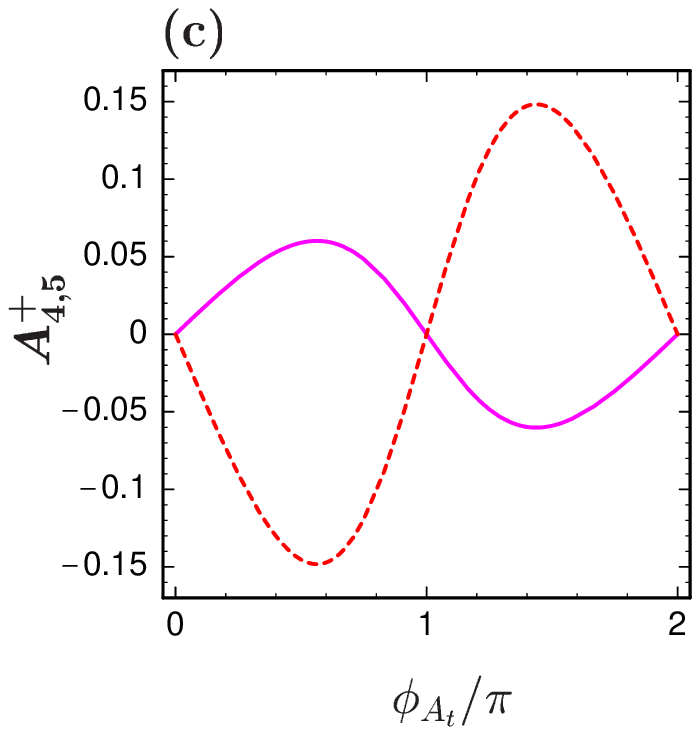,height=22.cm,width=19.4cm}}}
\put(27,-125){\mbox{\epsfig{figure=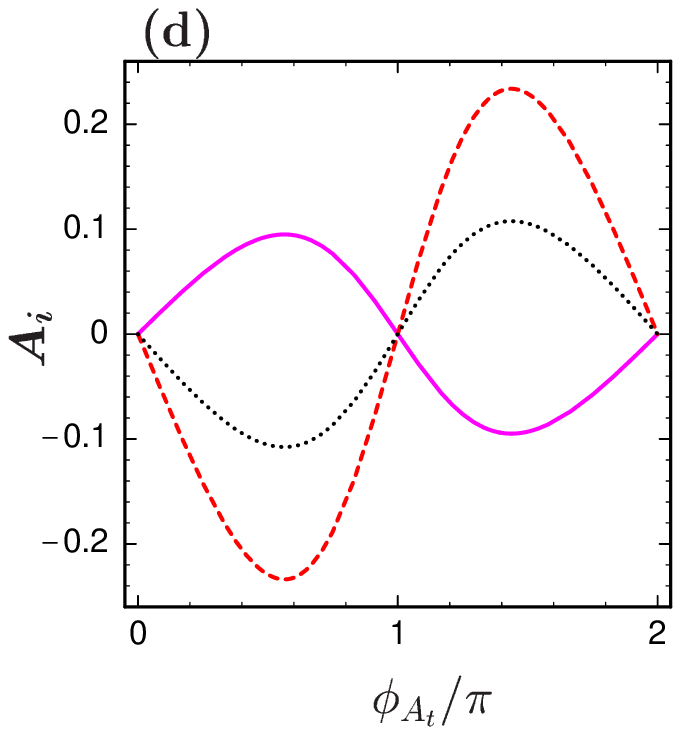,height=22.cm,width=19.4cm}}}
\end{picture}
\end{center}
\vspace{-2cm}
\caption{T-odd asymmetries (a) $A^+_i$, $i=1,2,3$, 
Eq.~\rf{eq:asym1}-\rf{eq:asym3},
(b) $A'^-_i$, $i=1,2,3$, Eq.~\rf{eq:asymp1}-\rf{eq:asymp3}, 
(c) $A^+_i$, $i=4,5$, Eqs.~\rf{eq:asym4} and \rf{eq:asym5},
and (d) $A_i$, $i=1,2,3$, Eqs.~\rf{eq:asymchar1}-\rf{eq:asymchar3}
for $\tilde{t}_1\to t\tilde{\chi}^0_2$ 
as a function of $\phi_{A_t}$. 
In (a), (b), (d) the solid (dashed, dotted) lines correspond to 
the indices $i=1~(2,3)$, in (c) the solid (dashed) line corresponds to
$i=4~(5)$.
The MSSM parameters 
are chosen as $|A_t|=1200$~GeV, $M_2=250$~GeV, $|\mu|=200$~GeV, 
$\tan\beta=10$, $\phi_{M_1}=\phi_{\mu}=0$, 
$m_{\ti t_1}=400$~GeV, $m_{\ti t_2}=800$~GeV, $M_{\ti Q}<M_{\ti U}$,
for $l=e,\mu$.}
\label{fig:fig1}
\end{figure}

\begin{figure}[t]
\setlength{\unitlength}{1mm}
\begin{center}
\begin{picture}(150,120)
\put(-53,-65){\mbox{\epsfig{figure=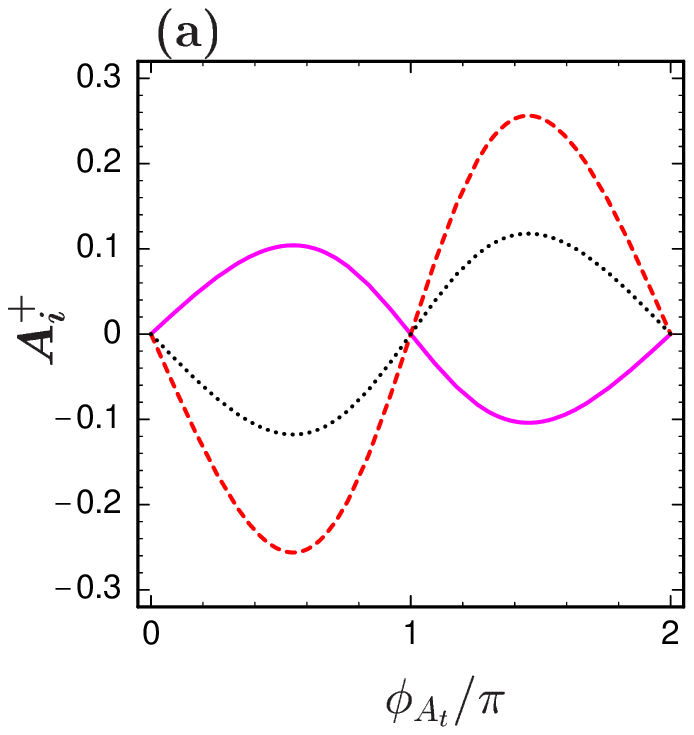,height=22.cm,width=19.4cm}}}
\put(27,-65){\mbox{\epsfig{figure=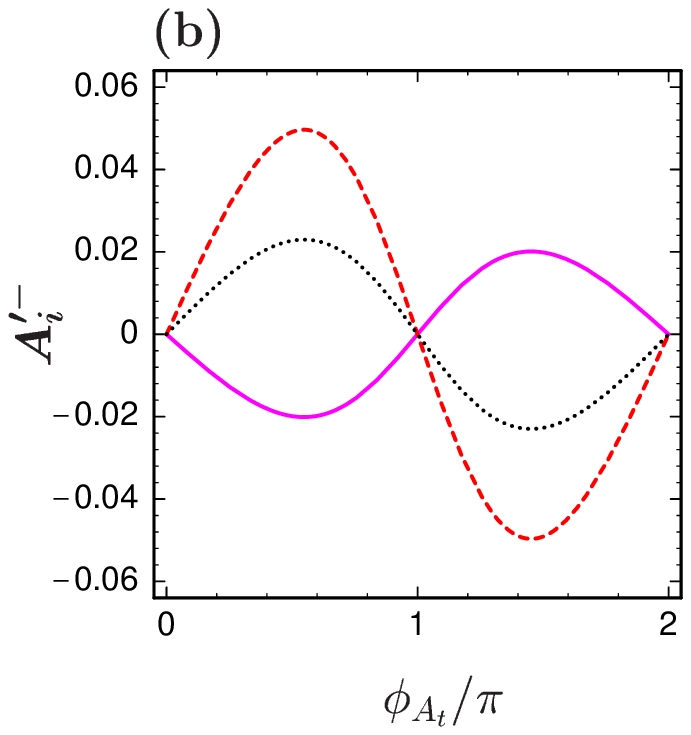,height=22.cm,width=19.4cm}}}
\put(-53,-125){\mbox{\epsfig{figure=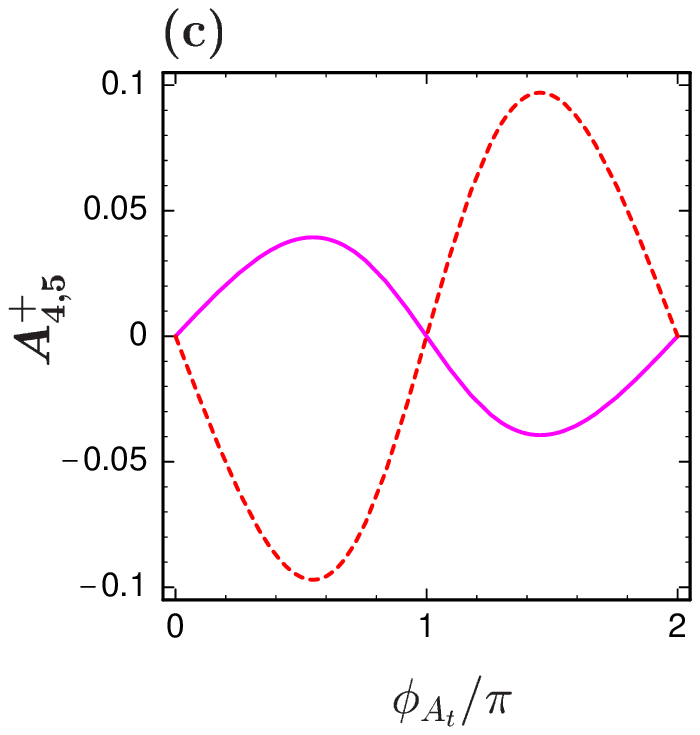,height=22.cm,width=19.4cm}}}
\put(27,-125){\mbox{\epsfig{figure=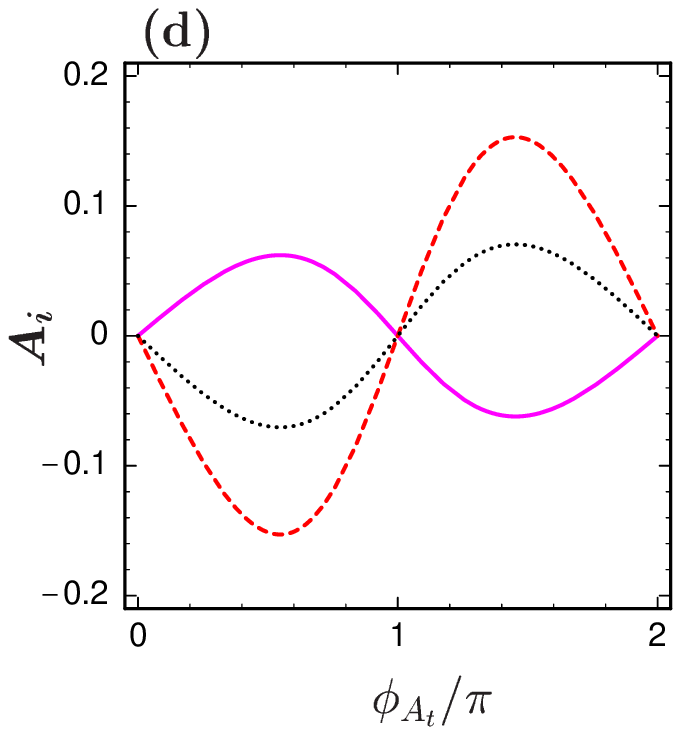,height=22.cm,width=19.4cm}}}
\end{picture}
\end{center}
\vspace{-2cm}
\caption{T-odd asymmetries (a) $A^+_i$, $i=1,2,3$, Eq.~\rf{eq:asym1}-\rf{eq:asym3},
(b) $A'^-_i$, $i=1,2,3$, Eq.~\rf{eq:asymp1}-\rf{eq:asymp3}, 
(c) $A^+_i$, $i=4,5$, Eqs.~\rf{eq:asym4} and \rf{eq:asym5},
and (d) $A_i$, $i=1,2,3$, Eqs.~\rf{eq:asymchar1}-\rf{eq:asymchar3}
for $\tilde{t}_1\to t\tilde{\chi}^0_2$ 
as a function of $\phi_{A_t}$.
In (a), (b), (d) the solid (dashed, dotted) lines correspond to 
the indices $i=1~(2,3)$, in (c) the solid (dashed) line corresponds to
$i=4~(5)$.
The MSSM parameters are chosen as $|A_t|=1200$~GeV, 
$M_2=250$~GeV, $|\mu|=200$~GeV, 
$\tan\beta = 10$, $\phi_{M_1}=\phi_{\mu}=0$, 
$m_{\ti t_1}=400$~GeV, $m_{\ti t_2}=800$~GeV, $M_{\ti Q}>M_{\ti U}$,
for $l=e,\mu$.}
\label{fig:fig2}
\end{figure}

In Fig.~\ref{fig:fig1} we plot the various asymmetries
\rf{eq:asym1}-\rf{eq:asymchar3} 
for the decay $\tilde{t}_1\to t\tilde{\chi}^0_2$
as a function of $\phi_{A_t}$ for the case $M_{\ti Q}<M_{\ti U}$.
The MSSM parameters are $M_2=250$~GeV, $|\mu|=200$~GeV,
$|A_t|=1200$~GeV and $\phi_{M_1}=\phi_{\mu}=0$. 
As can be seen in Fig.~\ref{fig:fig1}a the absolut value of the 
asymmetry $A^+_2$ (dashed line)
is much larger than the absolut value of $A^+_1$ (solid line),
which can be attributed to the sensitivity factor
of the top quark polarization for the two asymmetries.
For $A^+_1$ this factor is $|\alpha_b|\simeq 0.4$ 
(for $m_t=178$~GeV), Eq.~(\ref{eq:alphab}), whereas for $A^+_2$ it 
is $|\alpha_l|=1$, Eq.~\rf{eq:alphap} 
(see also \cite{Czarnecki:1990pe} where QCD corrections are included).
This difference can also be seen in Fig.~\ref{fig:fig1}c
by comparing $A^+_4$, Eq.~\rf{eq:asym4}, with $A^+_5$, Eq.~\rf{eq:asym5}.
In Fig.~\ref{fig:fig1}d the asymmetries $A_i~(i=1,2,3)$, 
Eqs.~\rf{eq:asymchar1}-\rf{eq:asymchar3}, are displayed
for which we have to distinguish the leptons 
in the decay chain $\ti\chi^0_2 \to \ti l_1^{\mp} l_1^{\pm} 
\to \ti\chi^0_1 l^{\mp}_2 l^{\pm}_1$ only by there charge.
It is interesting to note that the asymmetry $A_2$ can be as 
large as 24\%. In Fig.~\ref{fig:fig2} 
we plot the same asymmetries as in Fig.~\ref{fig:fig1}, 
but now for the case $M_{\ti Q}>M_{\ti U}$.
As can be seen also for this case the largest asymmetry
is $A^+_2$, Eq.~\rf{eq:asym2}, which is however somewhat reduced
compared to the case $M_{\ti Q}<M_{\ti U}$. 

\begin{figure}[t]
\setlength{\unitlength}{1mm}
\begin{center}
\begin{picture}(150,120)
\put(-53,-110){\mbox{\epsfig{figure=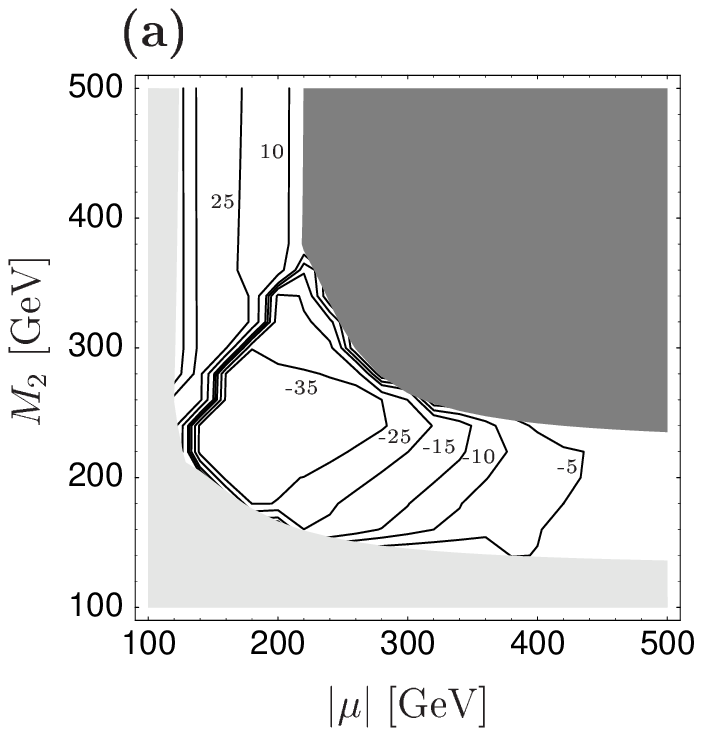,height=27cm,width=19.4cm}}}
\put(27,-110){\mbox{\epsfig{figure=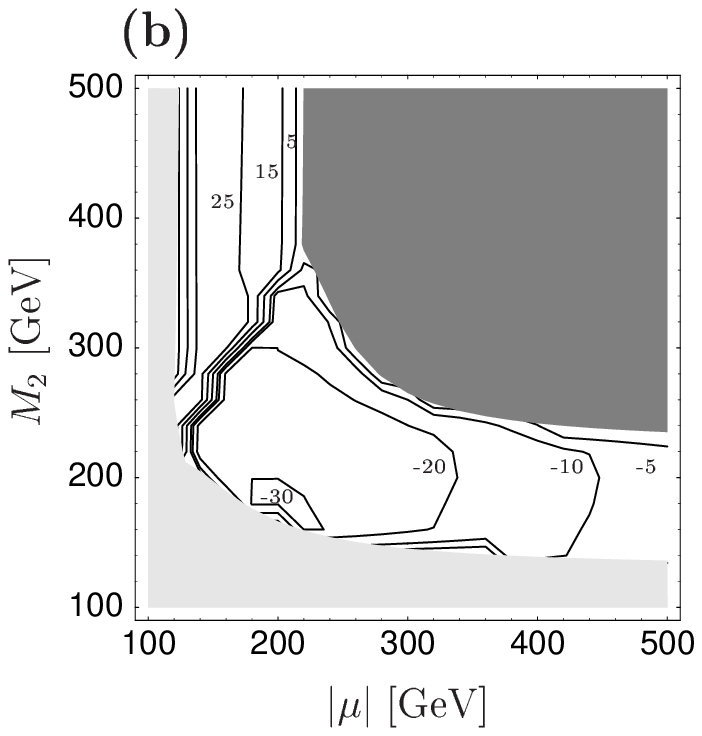,height=27cm,width=19.4cm}}}
\put(-53,-180){\mbox{\epsfig{figure=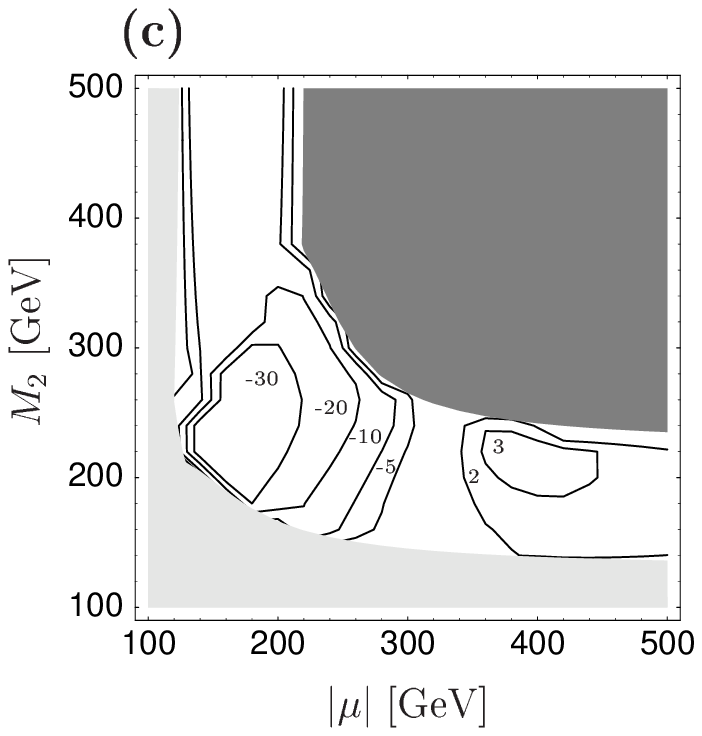,height=27cm,width=19.4cm}}}
\put(27,-180){\mbox{\epsfig{figure=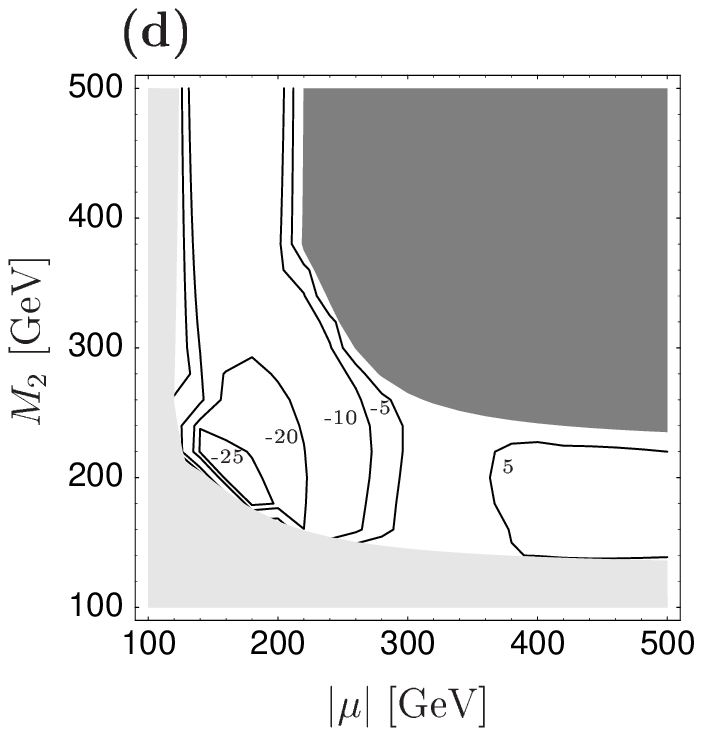,height=27cm,width=19.4cm}}}
\end{picture}
\end{center}
\caption{Contours of the T-odd asymmetry $A^+_2$ in \%
for $\tilde{t}_1\to t\tilde{\chi}^0_2\to bW^+\tilde{l}^-_1l^+_1$
for $l=e,\mu$,
(a) $M_{\ti Q}<M_{\ti U}$ and (b) $M_{\ti Q}>M_{\ti U}$
and for $l=\tau$
(c) $M_{\ti Q}<M_{\ti U}$ and (d) $M_{\ti Q}>M_{\ti U}$.
The MSSM parameters 
are chosen as $|A_t|=1200$~GeV, $\phi_{A_t}=0.5\pi$, 
$\tan\beta = 10$, $\phi_{M_1}=\phi_{\mu}=0$, 
$m_{\ti t_1}=400$~GeV, $m_{\ti t_2}=800$~GeV, $|A_{\tau}|=500$~GeV, $\phi_{A_{\tau}}=0$,
$m_{\tilde{\tau}_1}=130$~GeV and $m_{\tilde{\tau}_2}=300$~GeV.
The light gray region is
excluded because there $m_{\ti\chi_1^{\pm}}<103$~GeV and/or
$m_{\ti\chi^0_2}<m_{\ti l_1}$. 
In the dark gray area the two-body decay $\tilde{t}_1\to t\tilde{\chi}^0_2$ 
is kinematically forbidden.}
\label{fig:fig3}
\end{figure}

In Fig.~\ref{fig:fig3} we plot the contours of the asymmetry $A^+_2$
for the decay $\tilde{t}_1\to t\tilde{\chi}^0_2$
in the $|\mu|-M_2$ plane where we have taken $\phi_{A_t}=\frac{\pi}{2}$ 
and the other parameters as in the previous figures.
One sees in Figs.~\ref{fig:fig3}a-\ref{fig:fig3}d 
that the asymmetry is largest
for large gaugino-higgsino mixing ($|\mu|\sim M_2$). 
Figs.~\ref{fig:fig3}a and b correspond to the case where
$\tilde{\chi}^0_2$ decays into $l=e,\mu$, wheras Figs.~\ref{fig:fig3}c and d
correspond to the case $\tilde{\chi}^0_2\to \ti\tau_1 \tau^+$.
The asymmetries in Figs.~\ref{fig:fig3}a and b are larger 
than those in Figs.~\ref{fig:fig3}c and d, because 
of the effect of scalar tau mixing which leads to 
$|\alpha_+|< 1$ (see Eq.~\rf{eq:defQplus}), while for $l=e,\mu$, 
Figs.~\ref{fig:fig3}a and b, we have $|\alpha_+|= 1$.
Moreover, in Figs.~\ref{fig:fig3}a and b there is a sign change
of the asymmetries because of a sign change of the prefactor
${\Im m}(a_{12}^{\tilde t*}b_{12}^{\tilde t})$.
This sign change does not appear in Figs.~\ref{fig:fig3}c and d, 
because it is compensated by a simultaneous sign change
of $\alpha_+$, which occurs due to a level crossing
of the states $\ti\chi^0_2$ and $\ti\chi^0_3$.
In Figs.~\ref{fig:fig3}c and d there is a different sign change
of the asymmetries in the lower right part
of the $|\mu|-M_2$ plane because $\alpha_+$, Eq.~\rf{eq:defQplus}, 
changes sign there.

\begin{figure}[t]
\setlength{\unitlength}{1mm}
\begin{center}
\begin{picture}(150,120)
\put(-53,-110){\mbox{\epsfig{figure=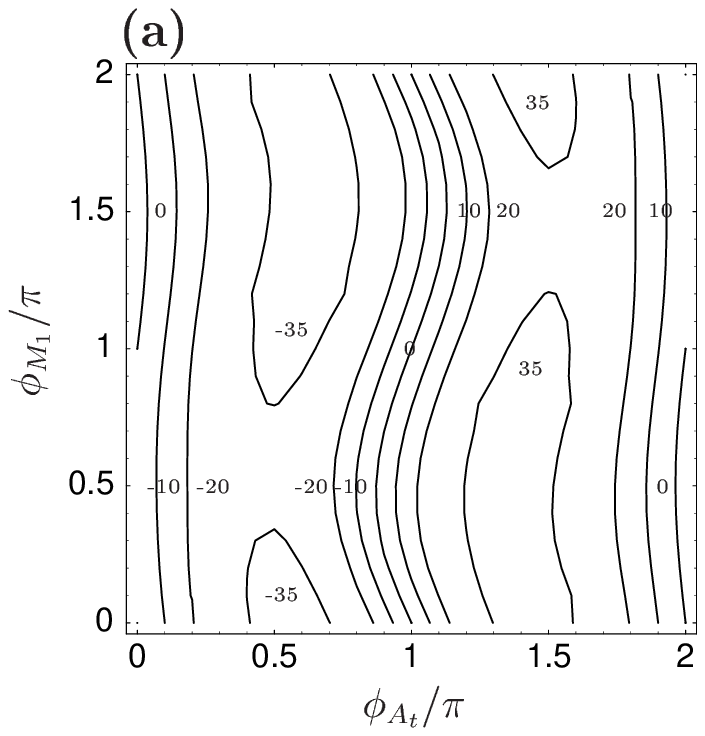,height=27cm,width=19.4cm}}}
\put(27,-110){\mbox{\epsfig{figure=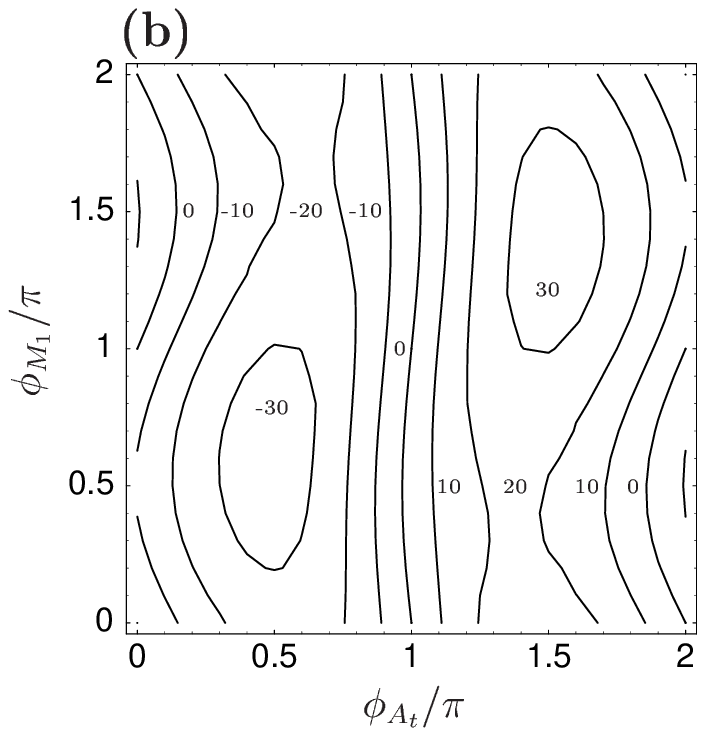,height=27cm,width=19.4cm}}}
\end{picture}
\end{center}
\vspace{-7cm}
\caption{Contours of the CP asymmetry $A^+_2$ in \%
for $\tilde{t}_1\to t\tilde{\chi}^0_2\to bW^+\tilde{l}^-_1l^+_1$. 
The MSSM parameters 
are chosen as $|A_t|=1200$~GeV, $M_2=250$~GeV, $|\mu|=200$~GeV,
$\phi_{\mu}=0$, $\tan\beta = 10$, 
$m_{\ti t_1}=400$~GeV, $m_{\ti t_2}=800$~GeV, for $l=e,\mu$,
(a) $M_{\ti Q}<M_{\ti U}$ and (b) $M_{\ti Q}>M_{\ti U}$.}
\label{fig:fig4}
\end{figure}

\begin{figure}
\setlength{\unitlength}{1mm}
\begin{center}
\begin{picture}(150,44)
\put(-53,-150){\mbox{\epsfig{figure=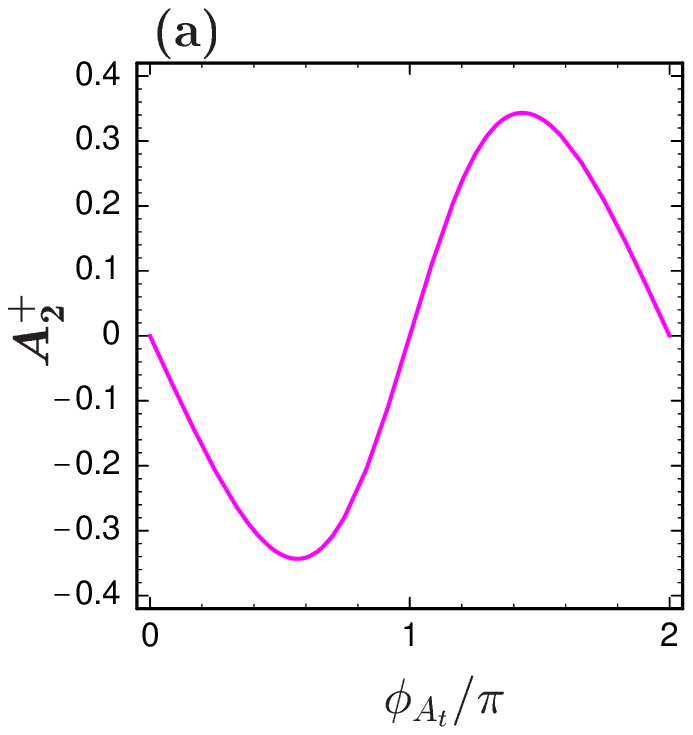,height=22.cm,width=19.4cm}}}
\put(27,-150){\mbox{\epsfig{figure=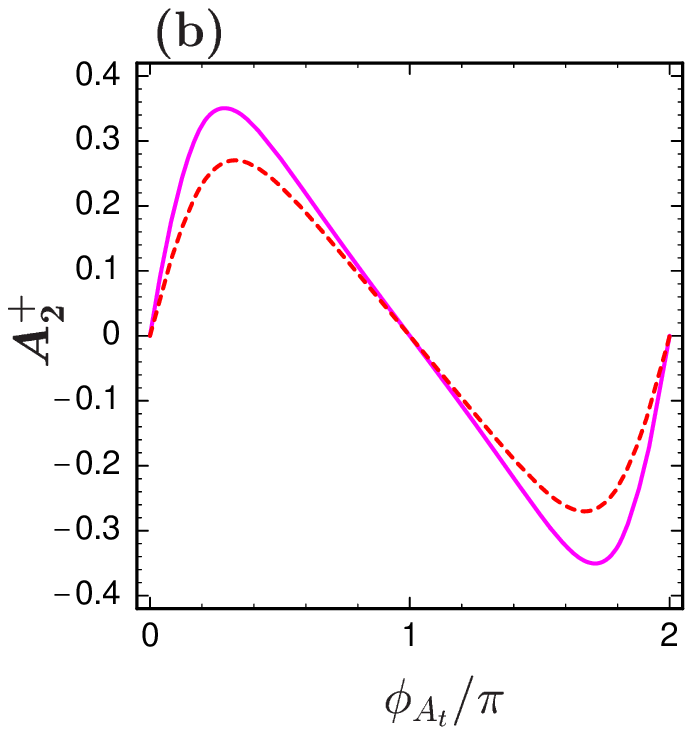,height=22.cm,width=19.4cm}}}
\end{picture}
\end{center}
\caption{T-odd asymmetry $A^+_2$, Eq.~\rf{eq:asym2},
for the processes (a) $\tilde{t}_2\to t\tilde{\chi}^0_3\to bW^+\tilde{l}^-_1l^+_1$ and (b) 
$\tilde{t}_2\to t\tilde{\chi}^0_4\to bW^+\tilde{l}^-_1l^+_1$ ($l=e,\mu$)
as a function of $\phi_{A_t}$. The MSSM parameters 
are chosen as $|A_t|=1200$~GeV, $M_2=|\mu|=200$~GeV, $
\tan\beta = 10$, $\phi_{M_1}=\phi_{\mu}=0$, 
$m_{\ti t_1}=400$~GeV, $m_{\ti t_2}=800$~GeV
for $M_{\ti Q}<M_{\ti U}$ (solid line) and $M_{\ti Q}>M_{\ti U}$ (dashed line).}
\label{fig:fig5}
\end{figure} 

In Fig.~\ref{fig:fig4} we show the contours of the asymmetry 
$A^+_2$ for the decay
$\tilde{t}_1 \to t\tilde{\chi}^0_2 \to bW^+\tilde{l}^-_1l^+_1~(l=e,\mu)$
in the $\phi_{A_t}-\phi_{M_1}$ plane taking $A_t=1200$~GeV,
$M_2=250$~GeV, $|\mu|=200$~GeV and $\phi_{\mu}=0$
for the two cases $M_{\tilde{Q}}< M_{\tilde{U}}$ (Fig.~\ref{fig:fig4}a)
and $M_{\tilde{Q}}> M_{\tilde{U}}$ (Fig.~\ref{fig:fig4}b).
In Fig.~\ref{fig:fig4}a the largest value of about $39\%$ ($-39\%$)
for the asymmetry $A^+_2$ is obtained for $\phi_{A_t}=1.4\pi$
($\phi_{A_t}=0.6\pi$).
For $\phi_{A_t}=0~(\pi)$ and $\phi_{M_1}=0.5\pi$ the asymmetry $A^+_2$ is
about $-3.5~(17.3)\%$.
In Fig.~\ref{fig:fig4}b one can see that the largest value of the asymmetry $A^+_2$
is obtained if $\phi_{M_1}\neq 0,\pi$ and the asymmetry
varies from about 25\% for $\phi_{A_t}\approx 1.5\pi,\phi_{M_1}=0$ to 
about 34\% for $\phi_{A_t}\approx 1.5\pi,\phi_{M_1}\approx 1.4 \pi$.
For $\phi_{A_t}=0~(\pi)$ and $\phi_{M_1}=0.5\pi$ the asymmetry $A^+_2$ 
is about $-10.6~(3.2)\%$.

In Fig.~\ref{fig:fig5} we show the asymmetry  $A^+_2$, Eq.~\rf{eq:asym2}, 
as a function of 
$\phi_{A_t}$ for the decay of the heavier top squark.
The MSSM parameters chosen are $|A_t|=1200$~GeV, $M_2=|\mu|=200$~GeV 
and $\phi_{M_1}=\phi_{\mu}=0$. 
Fig.~\ref{fig:fig5}a displays the asymmetry for the decay 
$\ti t_2\to t\tilde{\chi}^0_3$
and Fig.~\ref{fig:fig5}b shows the asymmetry for $\ti t_2\to t\tilde{\chi}^0_4$.
Figs.~\ref{fig:fig5}a and b demonstrate that also for the decay of $\ti t_2$
into the heavier neutralinos the asymmetry $A^+_2$ can be quite large.
In Fig.~\ref{fig:fig5}a the two cases $M_{\ti Q}<M_{\ti U}$ and $M_{\ti Q}>M_{\ti U}$ 
give nearly the same curve for $A^+_2$, because the value of
the decay width is nearly the same for the two cases.
Note that an observation of a T-odd asymmetry would lead to a two 
fold ambiguity in the extraction of the CP phases, which can be seen in 
Figs.~\ref{fig:fig1},\ref{fig:fig2},\ref{fig:fig4} and \ref{fig:fig5}.

Next we give a theoretical estimate of the number of
top squarks $\tilde{t}_1$ necessary to observe the
T-odd asymmetries \rf{eq:asym1}-\rf{eq:asymchar3} in the decay
$\ti t_1\to\ti\chi^0_2 t$, where subsequently the 
neutralino $\ti\chi^0_2$ decays into $\ti l_1 l$~ ($l=e,\mu,\tau$).
This number can be estimated by
\be{eq:estimate}
N_{\tilde{t}_1}\gsim
\frac{\sigma^2}{(A_T)^2 B(W\to f) B(\tilde{t}_1\to \tilde{\chi}^0_2 t)
B(\ti\chi^0_2\to \ti l_1 l)},
\ee
where $\sigma$ denotes the number of standard deviations and 
$A_T$ stands for any of the above asymmetries.
The value of the branching ratios of the $W$ boson is given by 
$B(W\to f)=(32,68,32)\% ~(f=\sum_l \nu l,
\sum_q q \bar q',c s)$ \cite{particledata} 
corresponding to the asymmetry for which the estimate is made.
For instance, for the asymmetry $A^{\pm}_1$, which is based
on the triple product $({\bf p}_b {\bf p}_t {\bf l}_1^{\pm})$, 
we take the value for $B(W \to \sum_q q \bar q')$.   
The estimate is taken for two scenarios which we define 
in Table~\ref{tab:scenarios}.
The scenarios chosen imply that $\sum_l B(\tilde{\chi}^0_2 \to \tilde{l}_1 l)=1~
(l=e,\mu,\tau)$ and that the energy distributions of the two final leptons
(from the decay $\ti\chi^0_k\to \ti l^-_1 l_1^+$ and from
the decay $\ti l^-_1\to \ti\chi^0_1 l_2^-$) do not overlap.
This means that in all decays it is possible
to decide from which decay the two leptons originate.
For simplicity we will assume that the T-odd asymmetries
\rf{eq:asym1}-\rf{eq:asymchar3} are equal for the three flavors
in the subsequent decay $\ti\chi^0_2\to \ti l_1 l$.
This means that we neglect scalar tau mixing ($|\alpha_+|=1$ for $l=e,\mu,\tau$)
and in addition we take
$m_{\ti l_1}=m_{\ti e_1}=m_{\ti \mu_1}=m_{\ti \tau_1}$.
For the calculation of the branching ratios of the $\ti t_1$ we use the
formulae given in \cite{Bartl:2003pd}. For scenario 1 (scenario 2) we obtain
$B(\tilde{t}_1\to \tilde{\chi}^0_2 t)=22 \%~ (2.1 \%)$ where
we assume that the bosonic decays $\ti t_1\to\ti b_1 W^+$ 
and $\ti t_1\to\ti b_1 H^+$ are kinematically not accessible.

\begin{table}[t]
\caption{The two scenarios used for the estimate of the 
necessary event rates.}
 \label{tab:scenarios}
 \begin{center}
 \begin{tabular}{|c|c|}
 \hline
scenario 1
&
scenario 2
\\
 \hline
 \hline
$m_{\ti l_1}=129$~GeV
  &
$m_{\ti l_1}=115$~GeV
  \\
$M_2=500$~GeV
  &
$M_2=200$~GeV
  \\
$|\mu|=150$~GeV
  &
$|\mu|=300$~GeV
  \\
$\tan\beta=3$
  &
$\tan\beta=6$
  \\
$\phi_{M_1}=\phi_{\mu}=0$
&
$\phi_{M_1}=0$, $\phi_{\mu}=\pi$
  \\
$|A_t|=1200$~GeV
&
$|A_t|=1200$~GeV
  \\
$\phi_{A_t}=\frac{\pi}{2}$
&
$\phi_{A_t}=\frac{\pi}{6}$
\\
$M_{\ti Q}<M_{\ti U}$
&
$M_{\ti Q}>M_{\ti U}$
\\
 \hline
 \end{tabular}
 \end{center}
\end{table}

\begin{table}[t]
\caption{The values of the T-odd asymmetries defined in
Eqs.~\rf{eq:asym1}-\rf{eq:asymchar3}
and the number $N_{\st_1}$ of top squarks
required to measure these asymmetries with a $3\sigma$ evidence
in the two considered scenarios (see Table~\ref{tab:scenarios}).}
 \label{tab:estimate}
 \begin{center}
 \begin{tabular}{|c||c|c||c|c|}
 \hline
 & \multicolumn{2}{|c||}{scenario 1}
 & \multicolumn{2}{|c|}{scenario 2} \\
 $A_T$ & value $[\%]$         & $N_{\ti t_1}\cdot 10^{-3} $
         & value $[\%]$         & $N_{\ti t_1}\cdot 10^{-3}$
           \\
 \hline
 $A^+_1$ &
 -11.5 &
 4.5 &
 15.9 &
 24.8
  \\
 $A^+_2$ &
 28.3 &
 1.6 &
 -39.0 &
 8.7 
  \\
 $A^+_3$ &
 13.8 &
 6.8 &
 -16.3 &
 50.4 
  \\
 \hline
 $A'^+_1$ &
 -1.3 &
 355.9 &
 5.1 &
 242.2
  \\
 $A'^+_2$ &
 3.2 &
 124.9 &
 -12.6 &
 84.3 
  \\
 $A'^+_3$ &
 1.6 &
 499.2 &
 -5.3 &
 476.8 
  \\
 \hline
 $A^+_4$ &
 -4.7 &
 18.5 &
 6.1 &
 115.0 
  \\
 $A^+_5$ &
 11.4 &
 9.8 &
 -14.9 &
 60.2 
  \\
 \hline
 $A_1$&
 -6.4 &
 14.8 &
 10.5 &
 57.1 
  \\
 $A_2$ &
 15.8 &
 5.2 &
 -25.8 &
 20.1 
  \\
 $A_3$ &
 7.7 &
 21.6 &
 -10.8 &
 114.8 
  \\
\hline
 \end{tabular}
 \end{center}
\end{table}

In Table \ref{tab:estimate} we display the values of the asymmetries,
Eqs.~\rf{eq:asym1}-\rf{eq:asymchar3}, and
the numbers $N_{\st_1}$ needed for a $3\sigma$ evidence of these asymmetries.
From Table~\ref{tab:estimate} it can be seen that in order to have
a $3\sigma$ evidence for
some of the T-odd asymmetries in scenario 1 about $O(10^3)$ produced $\ti t_1$'s
are necessary.
For scenario 2 $O(10^4)$ produced $\ti t_1$'s
are necessary for a $3\sigma$ evidence of some of the T-odd asymmetries.
Assuming that $O(10^6)$ $\ti t_1$'s can be produced
at the LHC and $O(10^5)$ $\ti t_1$'s at a future linear collider,
there are good prospects to measure some of the asymmetries. 
It is however clear that detailed Monte Carlo studies
taking into account background and detector simulation
are necessary to predict the expected accuracy. This is, however,
beyond the scope of the present paper.

\section{Summary and conclusion \label{sec:7}}

We have proposed a set of T-odd asymmetries in the decay 
$\ti t_m\to t \ti\chi_k^0$ with the subsequent decays $t\to bW^+\to bl\nu$ and
$\tilde\chi^0_k \to l^\pm\tilde l_n^\mp \to l^\pm l^\mp\tilde\chi^0_1$,
for $l =e,\mu,\tau$.
The asymmetries are based on triple product correlations 
involving the polarizations of the top quark and the $\ti\chi_k^0$ and arise
already at tree-level.
All the proposed T-odd asymmetries probe CP violation
in the $t-\ti t_m-\ti\chi_k^0$ couplings and are
proportional to the product of left- and right-couplings. 
Since top squark mixing is naturally
large due to the large top Yukawa coupling these asymmetries
may be large and will allow to determine the CP violating
phase $\phi_{A_t}$, which is not easily accessible otherwise.

In a numerical study of the T-odd asymmetries we have found that
the asymmetry $A^{\pm}_2$, which is based on the triple
product $({\bf p}_l {\bf p}_t {\bf l}_1^{\pm})$, is the largest 
one and its magnitude can go up to $40\%$, while the others are smaller.
We have also found that the asymmetry $A_2$, Eq.~\rf{eq:asymchar1}, based on 
$({\bf p}_l {\bf p}_t {\bf l}^{\pm})$, where
$l^\pm$ can be any of the final leptons $l^\pm_1$ and $l^\mp_2$, 
distinguished only
by their charges, is $\lsim 26\%$.
Moreover, we have made a theoretical estimate of the number
of $\ti t_1$ necessary to observe the T-odd asymmetries
for two scenarios. Depending on the MSSM parameters,
we have found that a $\ti t_1$ production rate of $O(10^3)$
may be sufficient to observe some of the proposed T-odd asymmetries, 
which could be possible at the LHC or
at a future linear collider.

\section*{Acknowledgments}

This work is supported by the `Fonds zur
F\"orderung der wissenschaftlichen Forschung' (FWF) of Austria, project
No. P16592-N02 and by the European Community's
Human Potential Programme
under contract HPRN--CT--2000--00149. The work of E.C. was supported by
the Bulgarian National Science Foundation, Grant Ph-1010.

\section*{Appendix}


\section*{A. Neutralino Masses and Mixing}
\label{app:neut}

At tree-level the neutralino mass matrix in the weak basis
$(\tilde B, \tilde W^3, \tilde H_1^0, \tilde H_2^0)$ is given as
\cite{susy,guha}:
\begin{equation}
\mathcal{M}_N = \ \left( \begin{array}{cccc}
|M_1| e^{i\phi_{M_1}} & 0 & -m_Z s_W c_{\beta} & m_Z
s_W s_{\beta}\\[3mm]
0 & M_2 & m_Z c_W c_{\beta} & -m_Z c_W s_{\beta}\\[3mm]
-m_Z s_W c_{\beta} & m_Z c_W c_{\beta} & 0 &
-|\mu| e^{i\phi_{\mu}}\\[3mm]
m_Z s_W s_{\beta} & -m_Z c_W s_{\beta} &
-|\mu| e^{i\phi_{\mu}} & 0
\end{array}\right),
\label{eq:massN}
\end{equation}
where $\phi_{M_1}$ is the phase of $M_1$, and
$c_W$ and $s_W$ are $\cos \theta_W$ and $\sin \theta_W$, respectively.
This symmetric complex mass matrix is diagonalized by the
unitary $4 \times 4$ matrix $N$:
\begin{equation}
\label{eq:mixN}
N^{\ast}\mathcal{M}_N N^{\dag}\ = \mathrm{diag}(m_{\tilde\chi_1^0},\dots,
m_{\tilde\chi_4^0}),
\hspace{2cm} 0\le m_{\tilde\chi_1^0} \le \dots \le m_{\tilde\chi_4^0} \,.
\end{equation}


\section*{B. Masses and mixing in squark sector}
\label{app:squarks}

The left-right mixing of the top squarks is described by a
hermitian $2 \times 2$ mass matrix which in the basis
$(\tilde{t}_L,\tilde{t}_R)$ reads
\be{eq:stopmass}
{\mathcal{L}}_M^{\st}= -(\st_L^{\dagger},\, \st_R^{\dagger})
\left(\begin{array}{ccc}
M_{\st_{LL}}^2 & e^{-i\phi_{\st}}|M_{\st_{LR}}^2|\\[5mm]
e^{i\phi_{\st}}|M_{\st_{LR}}^2| & M_{\st_{RR}}^2
\end{array}\right)\left(
\begin{array}{ccc}
\st_L\\[5mm]
\st_R \end{array}\right),
\ee
where
\begin{eqnarray}
M_{\st_{LL}}^2 & = & M_{\tilde Q}^2+(\frac{1}{2}-\frac{2}{3}\sin^2\Theta_W)
\cos2\beta \ m_Z^2+m_t^2 ,\label{eq:mll} \\[3mm]
M_{\st_{RR}}^2 & = & M_{\tilde U}^2+\frac{2}{3}\sin^2\Theta_W\cos2\beta \
m_Z^2+m_t^2 ,\label{eq:mrr}\\[3mm]
M_{\st_{RL}}^2 & = & (M_{\st_{LR}}^2)^{\ast}=
m_t(A_t-\mu^{\ast}
\cot\beta), \label{eq:mlr}
\end{eqnarray}
\begin{equation}
\phi_{\st}  = \arg\lbrack A_{\st}-\mu^{\ast}\cot\beta\rbrack ,
\label{eq:phtau}
\end{equation}
where $\tan\beta=v_2/v_1$ with $v_1 (v_2)$ being the vacuum
expectation value of the Higgs field $H_1^0 (H_2^0)$,
$m_t$ is the mass of the top quark and
$\Theta_W$ is the weak mixing angle, $\mu$ is the Higgs--higgsino mass parameter
and $M_{\ti Q}$,
$M_{\ti U}, A_t$ are the soft SUSY--breaking parameters of the top squark
system.
The mass eigenstates $\st_i$ are $(\ti t_1, \ti t_2)=
(\st_L, \st_R) {\mathcal{R}^{\st}}^T$ with
\begin{equation}
\mathcal{R}^{\st}=\left( \begin{array}{ccc}
e^{i\phi_{\st}}\cos\theta_{\st} &
\sin\theta_{\st}\\[5mm]
-\sin\theta_{\st} &
e^{-i\phi_{\st}}\cos\theta_{\st}
\end{array}\right),
\label{eq:rtau}
\end{equation}
with
\begin{equation}
\cos\theta_{\st}=\frac{-|M_{\st_{LR}}^2|}{\sqrt{|M_{\st _{LR}}^2|^2+
(m_{\st_1}^2-M_{\st_{LL}}^2)^2}},\quad
\sin\theta_{\st}=\frac{M_{\st_{LL}}^2-m_{\st_1}^2}
{\sqrt{|M_{\st_{LR}}^2|^2+(m_{\st_1}^2-M_{\st_{LL}}^2)^2}}.
\label{eq:thtau}
\end{equation}
The mass eigenvalues are
\begin{equation}
 m_{\st_{1,2}}^2 = \frac{1}{2}\left((M_{\st_{LL}}^2+M_{\st_{RR}}^2)\mp
\sqrt{(M_{\st_{LL}}^2 - M_{\st_{RR}}^2)^2 +4|M_{\st_{LR}}^2|^2}\right).
\label{eq:m12}
\end{equation}
Note here that for $|A_t|\gg |\mu| \cot\beta$ we have
$\phi_{\ti t}\approx\phi_{A_t}$.


\end{document}